%% file: TAU-24-002_temp.tex
\pdfoutput=1
\documentclass[11pt,twoside,a4paper,cmspaper,final,collab]{cms-tdr}
\def\svnVersion{51a7b6c}\def\svnDate{2026/04/02}\def\cmsCernNoTag{CERN-EP-2026-008}\def\cmsCernDate{\today}\def\cmsMessage{Published in the Journal of Instrumentation as \href{http://dx.doi.org/10.1088/1748-0221/21/04/P04002}{\doi{10.1088/1748-0221/21/04/P04002}.}}
\begin{document}\cmsNoteHeader{TAU-24-002}

\hyphenation{had-ron-i-za-tion}
\hyphenation{cal-or-i-me-ter}
\hyphenation{de-vices}

\newlength\cmsFigWidth
\setlength\cmsFigWidth{0.4\textwidth}

\cmsNoteHeader{TAU-24-002}
\title{High-level hadronic tau lepton triggers of the CMS experiment in proton-proton collisions at \texorpdfstring{$\sqrt{s} = 13.6\TeV$}{sqrt(s) = 13.6 TeV}}

\date{\today}

\abstract{The trigger system of the CMS detector is pivotal in the acquisition of data for physics measurements and searches. Studies of final states characterized by hadronic decays of tau leptons require the reconstruction and the identification of genuine tau leptons against quark- and gluon-initiated jets at the trigger level. This is a difficult task, particularly as improvements to the LHC have resulted in an increased number of interactions per bunch crossing in recent years. To address this challenge, a series of machine-learning algorithms with high identification efficiency and low computational cost have been incorporated into the high-level trigger for hadronically decaying tau leptons. In this paper, these developments and the trigger performance are summarized using data collected by the CMS experiment in proton-proton collisions at $\sqrt{s} = 13.6\TeV$ in 2022--2023, corresponding to an integrated luminosity of 62\fbinv.
}

\hypersetup{
pdfauthor={CMS Collaboration},
pdftitle={High-level hadronic tau lepton triggers of the CMS experiment in proton-proton collisions at sqrt(s) = 13.6 TeV},
pdfsubject={CMS},
pdfkeywords={High-level trigger, tau trigger, tau reconstruction, DeepTau}}

\maketitle

\newcommand{\dEta}{\ensuremath{\Delta\eta}\xspace}
\newcommand{\dPhi}{\ensuremath{\Delta\phi}\xspace}
\newcommand{\dEtaTimesdPhi}{\ensuremath{\Delta\eta{\times}\Delta\phi}\xspace}
\newcommand{\EtaDashPhi}{\ensuremath{\eta\text{--}\phi}\xspace}
\newcommand{\PtDashEta}{\ensuremath{\pt\text{--}\eta}\xspace}
\newcommand{\dR}{\ensuremath{{\Delta}R}\xspace}
\newcommand{\dz}{\ensuremath{d_z}\xspace}
\newcommand{\NPV}{\ensuremath{N_{\text{PV}}}\xspace}

\newcommand{\abseta}{\ensuremath{\abs{\eta}}\xspace}
\newcommand{\penEta}{\ensuremath{\eta}\xspace}
\newcommand{\penPhi}{\ensuremath{\phi}\xspace}

\newcommand{\Rsig}{\ensuremath{R_{\text{sig}}}\xspace}
\newcommand{\nTT}{\ensuremath{n_{\mathrm{TT}}}\xspace}
\newcommand{\chisq}{\ensuremath{\chi^{2}}\xspace}

\newcommand{\PWj}{\ensuremath{\PW{+}\text{jet}}\xspace}
\newcommand{\PHtoPGtPGt}{\ensuremath{\PH\to\PGt\PGt}\xspace}
\newcommand{\PZprtoPGtPGt}{\ensuremath{\PZpr\to\PGt\PGt}\xspace}
\newcommand{\PpPp}{\ensuremath{\Pp\Pp}\xspace}

\newcommand{\mutauh}{\ensuremath{\PGm\tauh}\xspace}
\newcommand{\etauh}{\ensuremath{\Pe\tauh}\xspace}
\newcommand{\tauhpj}{\ensuremath{\tauh{+}\text{jet}}\xspace}

\newcommand{\LtwoTauNNTag}{\textsc{L2TauNNTag}\xspace}
\newcommand{\DeepTau}{\textsc{DeepTau}\xspace}
\newcommand{\kHzns}{\ensuremath{\text{kHz}}} 
\newcommand{\Hzns}{\ensuremath{\text{Hz}}} 
\newcommand{\textplus}{\ensuremath{+}}

\section{Introduction}\label{sec:introduction}
The tau lepton (\PGt) is widely used in measurements and searches carried out using proton-proton (\PpPp) collision data recorded at the CERN LHC. The tau lepton's importance is most apparent in the observation and the measurements of the standard model (SM) Higgs boson decaying to a pair of tau leptons~\cite{Sirunyan:2017khh,Sirunyan:2017djm,ATLAS:2018ynr,ATLAS:2018uni,ATLAS:2020evk,CMS:2024jbe,CMS:2022kdx}, the precision measurements of the tau lepton anomalous magnetic moment~\cite{CMS:2024qjo,ATLAS:2022ryk}, and the measurements of the tau lepton polarization~\cite{CMS:2023mgq,ATLAS:2017xuc,ATLAS:2012tqa}. Moreover, final states including tau leptons are at the forefront of searches for beyond-the-SM (BSM) particles, processes, and couplings~\cite{CMS:2022goy,ATLAS:2020zms}.

Tau leptons are massive enough to decay hadronically, and are the heaviest leptons in the SM, possessing a mass of 1.777\GeV and mean lifetime of 290\unit{fs}~\cite{PDG2024}. Because of the short lifetime, tau leptons produced by \PpPp collisions in CMS are not detected directly and can only be reconstructed from their decay products. Tau leptons decay hadronically with a branching fraction of 64.8\%, resulting in one tau neutrino (\PGnGt) and two quarks. The quarks undergo hadronization, and the final object is reconstructed as a hadronic tau (\tauh) candidate. The remaining branching fraction comprises leptonic decays, either to an electron or a muon, and accompanying neutrinos.

In this paper, we describe for the first time the developments of machine-learning (ML) algorithms deployed into the high-level trigger (HLT) in the CMS experiment~\cite{CMS:2008xjf,CMS:2023gfb} for identifying \tauh candidates. The performance is evaluated using \PpPp collision data collected in 2022--2023 at a center-of-mass energy of 13.6\TeV, which we refer to as ``early Run~3'' or simply ``Run~3'' throughout the paper. Similarly, we use ``Run~2'' to refer to the data-taking period from 2016--2018. Run~3 is the first time that the HLT is equipped with ML algorithms for \tauh candidates. Because of these advancements and the instrumental upgrades~\cite{CMS:2023gfb}, the HLT \tauh candidates are captured with higher efficiency than in Run~2~\cite{CMS:2018jrd}, despite an aging detector and an increased number of additional \PpPp interactions per bunch crossing.

This paper is structured as follows: Section~\ref{sec:cms detector} introduces the CMS detector with emphasis on its two-tiered trigger system, followed by a description of the offline event reconstruction in Section~\ref{sec:event reco}. Sections~\ref{sec:samples} and~\ref{sec:taureco} describe the samples used for HLT development and performance evaluation, and \tauh candidate reconstruction at the HLT, respectively. Sections~\ref{sec:L2NN} and~\ref{sec:deeptau} focus on the development of ML algorithms for \tauh candidate identification at the HLT. Section~\ref{sec:performance} discusses the performance of \tauh candidate reconstruction in early Run~3 data, and a summary is given in Section~\ref{sec:summary}.

\section{The CMS detector}\label{sec:cms detector}
The CMS apparatus~\cite{CMS:2008xjf,CMS:2023gfb} is a multipurpose, nearly hermetic detector, designed to trigger on~\cite{CMS:2020cmk,CMS:2016ngn,CMS:2024aqx} and identify electrons, muons, photons, and (charged and neutral) hadrons~\cite{CMS:2020uim,CMS:2018rym,CMS:2014pgm}. Its central feature is a superconducting solenoid of 6\unit{m} internal diameter, providing a magnetic field of 3.8\unit{T}. A silicon pixel and strip tracker, a lead tungstate crystal electromagnetic calorimeter (ECAL) with silicon strip preshower detectors placed at the front faces of the two endcap calorimeters, and a brass and scintillator hadron calorimeter (HCAL), each composed of a barrel and two endcap sections, are installed inside the solenoid. Forward calorimeters extend the pseudorapidity (\penEta) coverage provided by the barrel and endcap detectors. Muons are detected in gas-ionization chambers embedded in the steel flux-return yoke outside the solenoid. A more detailed description of the CMS detector, together with a definition of the coordinate system and the relevant kinematic variables, is presented in Refs.~\cite{CMS:2008xjf,CMS:2023gfb}.

Before the start of the 2022 data-taking period, numerous upgrades and improvements to the subdetectors, readout electronics, trigger, data acquisition, software, and offline computing systems were implemented. These include new silicon photomultipliers and readout electronics for the HCAL that allow for a finer granularity and longitudinal segmentation~\cite{CMS:2012tda}, the replacement of the innermost layer of the silicon pixel detector~\cite{CMS-DP-2022-047}, the new hybrid CPU/GPU farm for the HLT~\cite{CMS-DP-2023-004}, and the rebuilt dedicated online luminosity monitors~\cite{CMSBRIL:2022qin,Karunarathna:2022tyd,Wanczyk:2022avs}. 

Events of interest are selected using a two-tiered trigger system. The first level (L1), composed of custom hardware processors, uses information from the calorimeters and muon detectors to select events at a rate of around 100\unit{kHz} within a fixed latency of 4\mus~\cite{CMS:2020cmk}. The second level, known as the HLT, consists of a farm of processors running a version of the full event reconstruction software optimized for fast processing, and reduces the event rate to a few\unit{kHz} before data storage~\cite{CMS:2016ngn,CMS:2024aqx}. An HLT ``path'' is a sequence of algorithms and selections that performs a simplified event reconstruction during data taking and targets a physics process. In Run~3, the full HLT menu consists of about 800 different HLT paths~\cite{CMS:2016ngn}. The term ``online'' is used to refer to quantities calculated and physics objects produced at the time of data taking, to determine if an event will be stored for later analysis. If an event is accepted by the HLT based on its online information, a more sophisticated ``offline'' reconstruction is performed and used for further analysis.

\section{Reconstruction of offline objects}\label{sec:event reco}
This Section describes the general reconstruction methods of physics objects used in the training and performance evaluation for offline studies.

Final-state particles, such as electrons, muons, and charged and neutral hadrons, are reconstructed by the particle-flow (PF) algorithm~\cite{CMS:2017yfk}, which uses an optimized combination of information from subdetectors of the CMS experiment. The energy of photons is obtained from the ECAL measurement. The energy of electrons is determined from a combination of the electron momentum at the primary interaction vertex from the tracker, the energy of the corresponding ECAL cluster, and the energy sum of all bremsstrahlung photons spatially compatible with originating from the electron track. The momentum of muons is obtained from the curvature of the corresponding track. The energy of charged hadrons is determined from a combination of their momentum measured in the tracker and the matching ECAL and HCAL energy deposits, corrected for the response function of the calorimeters to hadronic showers. Finally, the energy of neutral hadrons is obtained from the corresponding corrected ECAL and HCAL energies. These particles are then used to reconstruct higher-level physics objects, including jets, missing transverse momentum, and \tauh candidates. The primary vertex is taken to be the vertex corresponding to the hardest scattering in the event, evaluated using tracking information alone, as described in Ref.~\cite{Contardo:2020886}. 

For each event, hadronic jets are clustered from these reconstructed particles using the infrared and collinear safe anti-\kt algorithm~\cite{Cacciari:2008gp, Cacciari:2011ma} with a distance parameter of 0.4. The jet momentum is determined as the vectorial sum of all particle momenta in the jet, and is found from simulation to be, on average, within 5 to 10\% of the generator-level momentum over the whole transverse momentum (\pt) spectrum and detector acceptance. Additional \PpPp interactions within the same or nearby bunch crossings (pileup) can contribute additional tracks and calorimetric energy depositions to the jet momentum. To mitigate this effect, charged particles identified to be originating from pileup vertices are discarded and an area-based offset correction is applied during Run~2~\cite{CMS:2020ebo}, whereas the pileup-per-particle identification algorithm is used during Run~3~\cite{Bertolini:2014bba}. Jet energy corrections are derived from simulation to bring the measured response of jets to that of particle level jets on average. In situ measurements of the momentum balance in dijet, photon\textplus jet, $\PZ {+} \text{jet}$, and multijet events are used to account for any residual differences in the jet energy scale between data and simulation~\cite{CMS:2016lmd}. The jet energy resolution amounts typically to 15--20\% at 30\GeV, 10\% at 100\GeV, and 5\% at 1\TeV~\cite{CMS:2016lmd}. Additional selection criteria are applied to each jet to remove jets potentially dominated by anomalous contributions from various subdetector components or reconstruction failures~\cite{CMS:2020ebo}.

Electrons are measured in the range $\abseta < 2.5$. The reconstruction efficiency in data ranges from 88 to 98\% in the barrel region ($\abseta < 1.479$) and from 90 to 96\% in the endcaps ($1.479 < \abseta < 2.5$) for electrons in the \pt range between 10 and 100\GeV. The momentum resolution for electrons with $\pt \approx 45\GeV$ from $\PZ\to\Pe\Pe$ decays ranges from 1.7 to 4.5\%. It is generally better in the barrel region than in the endcaps, and also depends on the bremsstrahlung energy emitted by the electron as it traverses the tracker material before reaching the ECAL~\cite{CMS:2015xaf}. 

Muons are measured in the range $\abseta < 2.4$, with detection planes made using three technologies: drift tubes, cathode strip chambers, and resistive plate chambers. The single muon trigger efficiency exceeds 90\% over the full \penEta range, and the efficiency to reconstruct and identify muons is greater than 96\%. Matching muons to tracks measured in the silicon tracker results in a relative \pt resolution of 1\% in the barrel and 3\% in the endcaps for muons with $\pt < 100\GeV$. Measurements made with cosmic ray muons show that, in the central region of the detector, the \pt resolution is better than 7\% for muons with \pt up to 1\TeV~\cite{CMS:2018rym}.

The missing transverse momentum vector, \ptvecmiss, is computed as the negative vectorial sum of all the PF candidate \pt in an event, and its magnitude is denoted as \ptmiss~\cite{CMS:2019ctu}. The \ptvecmiss is modified to include corrections to the energy scale of the reconstructed jets in the event.

Jets are used as seeds in the reconstruction of \tauh candidates, which is carried out with the hadron-plus-strips (HPS) algorithm~\cite{CMS:2011eio,CMS:2015pac,CMS:2018jrd}, analogously to the logic in the HLT described in Section~\ref{sec:taureco}. To distinguish genuine \tauh decays from jets originating from the hadronization of quarks or gluons, and from electrons or muons, the \DeepTau algorithm is used~\cite{DeepTau_CMS-DP-2019-033,CMS:2022prd}, and its online implementation is discussed in detail in Section~\ref{sec:deeptau}.

\section{Data and simulated samples}\label{sec:samples}
The early Run~3 data collected by the CMS detector correspond to 34.7 (27.2)\fbinv of \PpPp collisions from 2022 (2023)~\cite{CMS:LUM-22-001,CMS-DP-2024-068,CMS:LUM-21-001,CMS:2021xjt}. Further event selection criteria are applied depending on the analyses performed and are described in detail in the relevant Sections.

Monte Carlo (MC) samples are produced for ML algorithm training and trigger performance evaluation. Drell--Yan (DY) ($\PZ/\PGg^{*}\to\Pell\Pell$, where $\Pell = \Pe, \PGm, \PGt$) and \PWj samples of simulated events are produced at leading order (LO) using \MGvATNLO~2.6.5~\cite{Alwall:2014hca} with the MLM jet merging scheme~\cite{Alwall:2007fs}. Events comprised predominantly of jets produced through the strong interaction, referred to as quantum chromodynamic (QCD) multijet events, are simulated at LO with \PYTHIA$\textsc{8.243}$~\cite{Sjostrand:2014zea}. Top quark-antiquark pairs (\ttbar) are simulated at next-to-LO with \POWHEG $\textsc{v2}$~\cite{Nason:2004rx,Frixione:2007vw,Alioli:2010xd,Campbell:2014kua,Frixione:2007nw}; this generator is also used for Higgs boson production via vector boson fusion at next-to-LO~\cite{Nason:2009ai} followed by the Higgs boson decaying to two tau leptons (\PHtoPGtPGt). The \PYTHIA generators are used together to produce a BSM sample of a heavy gauge boson decaying to two tau leptons (\PZprtoPGtPGt), where the \PZpr boson mass is 4\TeV.
The \PYTHIA generator, with the CP5 underlying-event tune~\cite{CMS:2019csb}, is interfaced with the matrix-element generators to model tau lepton decays by \TAUOLA~\cite{Davidson:2010rw}, the parton shower, and hadronization processes. The NNPDF 3.0 or 3.1 parton distribution functions are used as input in all the calculations~\cite{NNPDF:2014otw,NNPDF:2017mvq}. For all MC samples, the detector response is simulated using a detailed description of the CMS detector based on \GEANTfour~\cite{GEANT4:2002zbu}, and event reconstruction is performed on simulation with the same algorithms as are used for data. 

The simulated events are reweighted with the same pileup profile as found in the observed data. In 2022 and 2023,
the average number of interactions per bunch crossing was 46 and 52, respectively~\cite{CMS:LUM-22-001}.

\section{\texorpdfstring{Reconstruction of \tauh candidates at the HLT}{Reconstruction of hadronic tau candidates in the HLT}}\label{sec:taureco}
\subsection{Tau trigger reconstruction}
The tau lepton decay branching fractions are summarized in Table~\ref{tab:tauDecayMode}. As described in the introduction, the tau lepton can decay leptonically or hadronically. The key property of hadronic tau lepton decays is that they primarily proceed through mesonic resonances via an electroweak charged current interaction. These hadronic decay products provide characteristic information to reconstruct and identify tau leptons in an event.

\begin{table}[!ht]
  \centering
	\topcaption{Decay modes and branching fractions ($\mathcal{B}$) of the tau lepton alongside the mesonic resonances primarily involved in hadronic tau lepton decays~\cite{PDG2024}.}
	\label{tab:tauDecayMode}
  \begin{tabular}{l c c}
	Decay mode & $\mathcal{B}~[\%]$ & Resonance [\MeVns{}] \\
	\hline  
	Leptonic decays \qquad 35.2 & & \\
	    \qquad $\PGtm \to \Pem \PAGne\PGnGt$ & 17.8 & \\
	    \qquad $\PGtm \to \PGmm \PAGnGm \PGnGt$ & 17.4 & \\ 
	Hadronic decays \qquad 64.8 & & \\
	    \qquad $\PGtm \to \ensuremath{\text{h}^{-}} \PGnGt$ & 11.5 & \\
	    \qquad $\PGtm \to \ensuremath{\text{h}^{-}} \PGpz \PGnGt$ & 26.0 & \PGrP{770} \\
	    \qquad $\PGtm \to \ensuremath{\text{h}^{-}} \PGpz \PGpz \PGnGt$ & 9.5 & \PaDoP{1260} \\
	    \qquad $\PGtm \to \ensuremath{\text{h}^{-}\text{h}^{-}\text{h}^{+}} \PGnGt$ & 9.8 & \PaDoP{1260} \\
	    \qquad $\PGtm \to \ensuremath{\text{h}^{-}\text{h}^{-}\text{h}^{+}} \PGpz \PGnGt$ & 4.8 & \\
	    \qquad Other hadronic modes & 3.2 & \\
  \end{tabular}
\end{table}

All HLT paths begin with, or are ``seeded'' by, an L1 trigger. The L1 \tauh candidate algorithm begins by identifying locally maximal energy deposits above a threshold of 2\GeV. Using only position and energy information from the ECAL and HCAL, the algorithm builds clusters and merges them based on proximity conditions determined by the size of ECAL and HCAL detector elements, and the position of the local maxima in \penEta. Then, the \tauh candidate's position is computed as an energy-weighted average centered around the largest energy deposit. This is an effective first step to identify \tauh candidates because hadronically decaying tau leptons typically have less hadronic activity than QCD-induced jets, leading to a smaller total size and fewer clusters of energy deposition. Isolation requirements to suppress the surrounding energy deposit are applied as a function of the \tauh candidate's total energy, \penEta, and the number of calorimeter trigger towers (\nTT) in order to suppress the QCD-induced jets, which are usually broader. Calorimeter trigger towers group $5{\times}5$ ECAL crystals in the barrel and the corresponding HCAL tower behind them into one object with a total size of $\dEtaTimesdPhi = 0.087{\times}0.087$, where \penPhi is the azimuthal angle in radians. The isolation requirements are loosened for high \nTT and high total energy to minimize the efficiency dependence on pileup and maximize the efficiency. A full description of all L1 trigger objects and algorithms is given in Ref.~\cite{CMS:2020cmk}. The efficiency of both isolated and nonisolated \tauh candidates from L1 is typically greater than 80\% for 10\GeV above its threshold value and greater than 90\% for 20\GeV above it~\cite{CMS-DP-2023-008}.

The HLT is designed to use fast and simple algorithms to quickly reject lower quality events before running more complex reconstruction on passing events. Consequently, the \tauh candidate reconstruction at the HLT is performed in several steps depending on the final state targeted by the path, which includes di-\tauh, single-\tauh, \etauh, and \mutauh. A brief overview of the Run~2 \tauh candidate reconstruction steps at the HLT, discussed extensively in Ref.~\cite{CMS:2024aqx}, is given first to contrast the updated Run~3 approach.

\begin{figure*}[hbtp!]
   \centering
     {\includegraphics[width=0.75\textwidth]{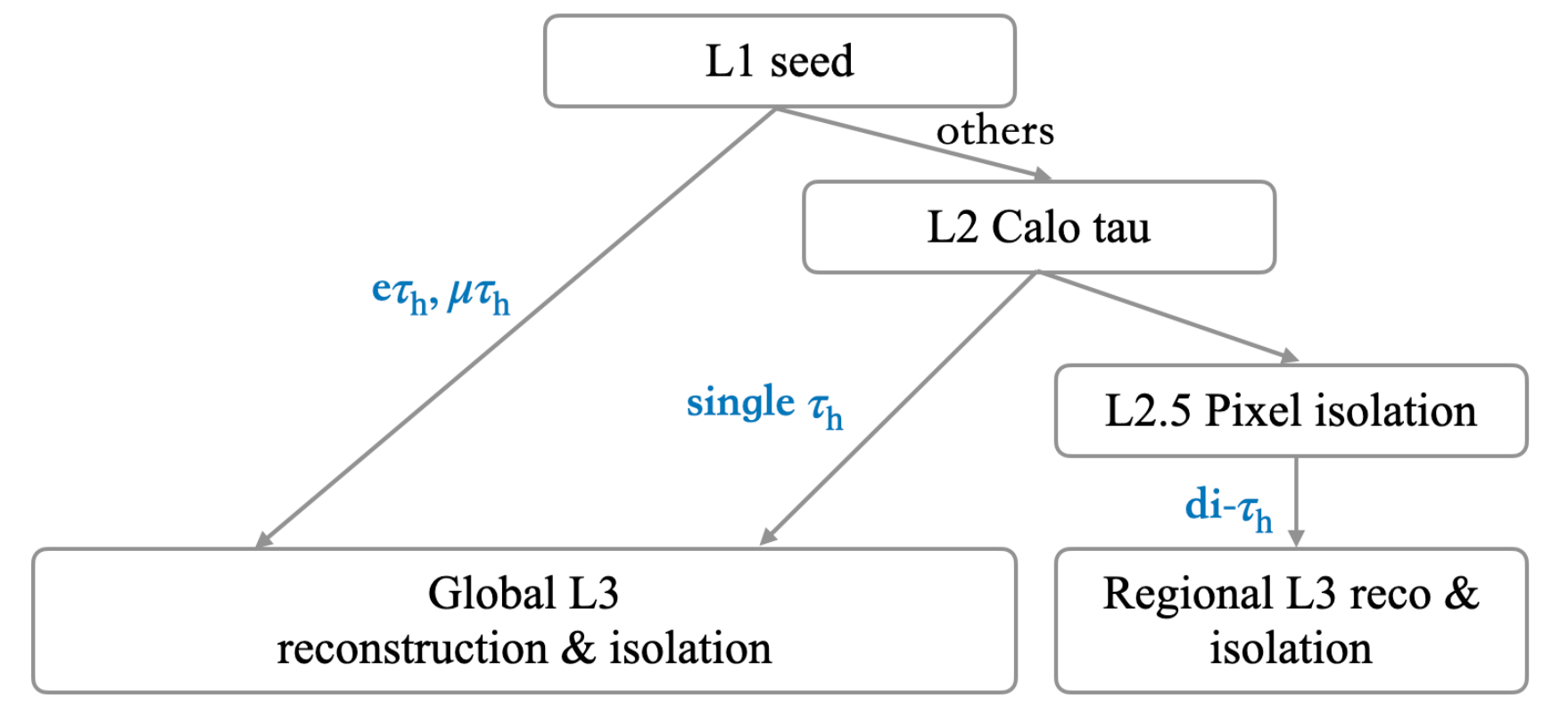}}
     \caption{Workflows for \tauh candidate reconstruction at the HLT in Run~2~\cite{CMS:2024aqx}.}\label{fig:workflowOLD}
\end{figure*}

In Run~2, the online workflow begins with L1 \tauh objects as seeds, and, depending on the HLT path, is then followed by so-called L2, L2.5, and L3 steps, which are progressively more complex. A diagram of the workflow is shown in Fig.~\ref{fig:workflowOLD}, as some HLT paths only need to run a subset of these steps.
The so-called cross trigger \etauh and \mutauh HLT paths already have low event rates due to the light-lepton reconstruction and identification, hence the L2 and L2.5 filtering steps are not necessary. For the single-\tauh and di-\tauh HLT paths, the L2 step combines L1 information with energy deposition information in the calorimeter towers, then the L2.5 step uses additional information from the pixel detector to compute a charged-particle isolation score. Finally, if the \tauh candidates pass the previous steps, they proceed to the L3 step, where online PF reconstruction is performed using the full tracking information. All paths except the di-\tauh one adopt a global reconstruction of all tracks in the event. The di-\tauh path applies a regional reconstruction with tracks localized around the L2 \tauh candidate. The algorithms used in Run~2 paths are referred to as ``cut-based'' algorithms because they enforce threshold values on relatively simple observables.

\begin{figure*}[hbtp!]
   \centering
     {\includegraphics[width=0.75\textwidth]{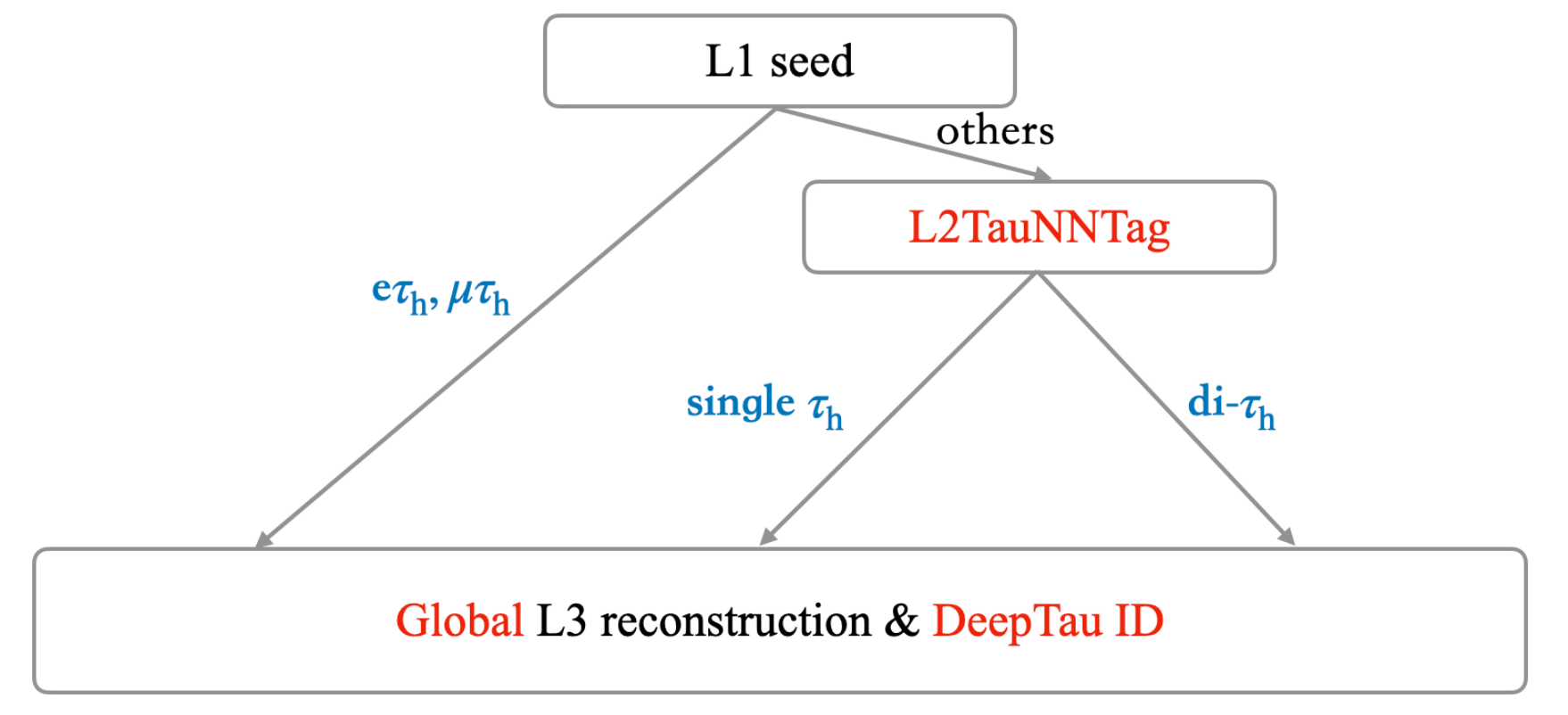}}
     \caption{Workflows for \tauh candidate reconstruction at the HLT in Run~3, since 2022.}\label{fig:workflowNEW}
\end{figure*}

In Run~3, the overall workflow is the same, but the cut-based algorithms previously used at the L2 and L3 steps have been replaced with ML algorithms, as shown in red in Fig.~\ref{fig:workflowNEW}. These updates include a convolutional neural network based algorithm at the L2 step, called \LtwoTauNNTag, that was developed to suppress the event rate to an affordable value in single-\tauh and di-\tauh HLT paths, as discussed in detail in Section~\ref{sec:L2NN}. The updated L3 reconstruction step adopts online \tauh candidate identification using a simplified version of the \DeepTau network, as discussed in Section~\ref{sec:deeptau}.

The \etauh and \mutauh cross triggers are designed to capture events with one \tauh candidate in association with an electron or muon. The \etauh HLT path is seeded with an electromagnetic object and a \tauh candidate from the L1 trigger.
The electromagnetic object can be an electron or photon, and is required to have $\pt > 22\GeV$ and satisfy a loose isolation criterion, determined by the number of hits in nearby trigger towers as a function of \pt. The L1 \tauh candidate is required to have $\pt > 26\GeV$ and satisfy similarly defined isolation criteria. The objects are required to not overlap within a cone of $\dR = \sqrt{\smash[b](\dEta)^2+(\dPhi)^2} < 0.3$. The \mutauh HLT path is seeded with an L1 muon object with $\pt > 18\GeV$ and an L1 \tauh candidate with $\pt > 24\GeV$. No isolation is required in the case of the \mutauh L1 trigger. 
At HLT, the electron (muon) is required to have $\pt > 24~(20)\GeV$ and satisfy tight identification and isolation criteria, which reduces the rate of events for the \tauh candidate reconstruction.

The L3 \tauh candidate reconstruction uses the full tracking information with the online PF reconstruction, and is a simplified version of the offline algorithm~\cite{CMS:2017yfk} in order to reduce the processing time.
With these inputs, a simplified version of the HPS algorithm is applied, modified from the offline \tauh candidate reconstruction, and discussed extensively in Ref.~\cite{CMS:2018jrd}. More details of the online HPS algorithm are given in Section~\ref{sec:HPS}.
The tau lepton identification (tau-ID) score of the reconstructed \tauh candidate is computed with the online version of \DeepTau. Finally, the \tauh candidate is required to have $\pt > 30~(27)\GeV$ for the \etauh (\mutauh) HLT path, and not overlap with the light lepton within a cone of $\dR < 0.5$.

The single-\tauh HLT path is designed to capture events with at least one high-\pt \tauh candidate. The HLT path is seeded with a \tauh candidate from L1 with $\pt > 120\GeV$, and no isolation requirement. The single-\tauh HLT path makes use of the \LtwoTauNNTag algorithm to reduce the rate of events entering the final L3 reconstruction. The previously described L3 reconstruction is then followed by the online tau-ID with \DeepTau, and finally a requirement that the \tauh candidate has $\pt > 180\GeV$.

The di-\tauh HLT path is meant to capture events with at least two \tauh candidates. It is seeded by two isolated L1 \tauh candidates, each with $\pt > 32\GeV$. This threshold was increased to 34\GeV in the later portion of 2022 data taking to mitigate a larger-than-expected L1 rate. If there are two L1 \tauh candidates with $\pt > 70\GeV$, no isolation criteria are applied. Similar to the single-\tauh HLT path, the \LtwoTauNNTag is used to reduce the input rate to the next reconstruction step. Again, the L3 reconstruction is performed, an online tau-ID score is evaluated with \DeepTau, and a \pt threshold of 35\GeV is applied to both \tauh candidates. Additionally, the candidates are required to not overlap within a cone of $\dR < 0.5$.

\subsection{The HPS algorithm}\label{sec:HPS}
The HPS algorithm was first used for the offline reconstruction of hadronic tau lepton decays in Run~2~\cite{CMS:2011eio,CMS:2015pac, CMS:2011eio,CMS:2015pac,CMS:2018jrd}. It was then introduced online to HLT paths with \tauh candidates in 2018 to replace the cone-based algorithm~\cite{CMS:2018jrd}. The initial inputs to the algorithm are PF jets with $\pt > 14\GeV$ and $\abseta < 2.5$ that were reconstructed using the anti-\kt algorithm with a distance parameter of $R=0.4$~\cite{CMS:2017yfk}. 

Targeting the hadronic decay modes in Table~\ref{tab:tauDecayMode}, the algorithm reconstructs \tauh candidates using the features of charged hadrons and neutral pions.
The neutral pions rapidly decay into photon pairs, a fraction of which can interact with the nuclei and associated electrons in the tracker material leading to photon conversion into $\Pep\Pem$ pairs. The strong magnetic field of the CMS detector's solenoid bends the pair of \Pep and \Pem into opposite directions in \penPhi, resulting in a spatial separation in the \EtaDashPhi plane. Photon and electron candidates that are found in the online PF jets are clustered into a ``strip'' with a \dEtaTimesdPhi area of $0.05{\times}0.2$ around the highest \pt photon or electron. The wider spread of electrons in the \penPhi direction is caused by the CMS experiment's solenoid. The momentum of the strip is defined as the vectorial sum of the momentum of its constituents, and the mass is assigned to be the \PGpz mass. Only the strips with $\pt > 2.5\GeV$ are considered. 

The cone size for reconstructing a \tauh candidate is defined as $\Rsig = 3\GeV/\pt$, with the limits of $0.05 < \Rsig < 0.10$, where \pt is that of the hadronic system. All strips and charged hadrons in the cone are used to reconstruct the \tauh candidate. It is possible to construct multiple \tauh candidates for a single jet, in which case the one with the largest \pt is selected.

The number of charged and neutral hadrons, $N_{\mathrm{h}^{\pm}}$ and $N_{\PGpz}$, are used to classify the reconstructed \tauh candidates. For each charged hadron, there is a distinguishable track, and the decay modes are subsequently grouped into ``\textit{n-prong}'' (the number of charged hadrons) cases. For 3-prong decay modes, the summed charge of the constituents is required to equal ${\pm} 1$. Since the 2-prong cases violate charge conservation, they are considered to be 3-prong decays where one charged hadron was misreconstructed or escaped the detector acceptance. Further requirements of matching with the resonances listed in Table~\ref{tab:tauDecayMode} are realized by constraining the invariant mass of charged and neutral particles to the mass windows of \PGrP{770} or \PaDoP{1260}, optimized under dedicated online conditions.

\section{\texorpdfstring{The \LtwoTauNNTag algorithm at the HLT}{The L2TauNNTag algorithm at the HLT}}\label{sec:L2NN}
In Run~3, the cut-based L2 and L2.5 sequences from Run~2 have been replaced by a convolutional neural network, \LtwoTauNNTag, described here for the first time.
This new algorithm adopts ML to accommodate more available information in order to improve the efficiency of genuine \tauh candidates in \tauh HLT paths, while lowering or maintaining the rate budget with respect to Run~2.
However, the event rates in \tauh HLT paths are still dominated by background from QCD jets.
The algorithm takes advantage of new GPU-based tracks from the pixel detector~\cite{Bocci:2020pmi,CMS-DP-2022-014} in addition to the calorimeter information.
Its training is performed on \PGt-enriched MC samples of DY, \ttbar, and \PWj for genuine hadronic tau lepton decays, as well as QCD events for jets misidentified as hadronic tau lepton decays. The generator-level \pt distribution of each sample is reweighted to obtain a uniform yield in each bin.

\subsection{Network architecture}
There are several types of input features used in the \LtwoTauNNTag, which will be listed in increasing order of granularity.
At the global level of an event, the number of vertices is reconstructed using \textsc{Patatracks}, a set of GPU-based software developed for optimizing and accelerating data processing for pixel track reconstruction in the HLT and offline computing~\cite{Bocci:2020pmi,CMS-DP-2022-014}.
At the object level, the properties of L1 \tauh candidates are used, including \pt, \penEta, \penPhi, and isolation.
Further down to the detector level, the energy deposit information in the ECAL and HCAL linked to the L1 \tauh candidates is also used, consisting of the total energy detected, the number of energy deposits, the distances in \penEta and \penPhi between these deposits and the \tauh candidate, a chi-squared (\chisq) value assessing the consistency of the detected energy with respect to the expected pattern of genuine tau leptons, as well as the total energy sum and energy deposit multiplicity for cases with nonzero \chisq values.
In addition to the calorimeter information, the \textsc{Patatrack} observables for each L1 \tauh candidate are used, including the number of associated tracks and the scalar sum of their \pt, a flag indicating the presence of an associated reconstructed vertex, the charge associated with the tracks, the \penEta and \penPhi distance between the tracks and the L1 \tauh candidate, the \chisq and number of degrees of freedom of the track reconstruction, as well as the longitudinal (\dz) and transverse impact parameters.

To structure the data, the input features are organized in a $5{\times}5$ grid in the \EtaDashPhi plane, and are connected to four convolution layers with a $1{\times}1$ window (numbers of filters are 80, 60, 40, and 20), followed by four convolution layers with a $2{\times}2$ window (numbers of filters are 20, 20, 20, and 40). These are then interfaced to three dense layers (numbers of nodes are 40, 40, and 20) and a final dense layer with a sigmoid function. Batch normalization and the rectified linear unit activation function are used in each layer before the final layer. The loss function used is binary cross entropy, and the total number of trainable parameters is 23\,701. The training procedure employs the Adam optimizer~\cite{Kingma:2014vow} with a learning rate of 0.001. Once the validation loss reaches its minimum value, the model parameters are stored. An early stopping criteria is applied, such that the training is terminated if the validation loss does not exhibit any improvement for ten consecutive epochs. The dataset is partitioned into three subsets, using 60\% of the total dataset for training, 20\% for testing, and 20\% for validation.

\subsection{\texorpdfstring{Performance of the \LtwoTauNNTag algorithm}{Performance of the L2TauNNTag algorithm}}
\label{sec:perf-l2nn}
To compare the computing performance of the \LtwoTauNNTag algorithm and the previous L2+L2.5 cut-based approach, the processing time and event rates were evaluated on a machine purpose-built to emulate the real conditions of the HLT computing farm. 
Three different datasets were used and linearly scaled by instantaneous luminosity to compare event rates from the cut-based and \LtwoTauNNTag algorithms, including Run~2 data, Run~2 data re-emulated under Run~3 trigger conditions, and Run~3 data, as described in Table~\ref{tab:L2TauNNTag_comparison_tauhtauh_path}.
In the di-\tauh HLT path, the \LtwoTauNNTag step results in reduced event rate to the next step of the HLT, with a similar processing time and improved efficiency compared to the L2+L2.5 cut-based approach. 

\begin{table}[htbp]
   \centering
	\topcaption{Rate estimation and observation for the cut-based and \LtwoTauNNTag algorithms. Column A scales Run~2 data collected by the cut-based algorithm to Run~3 conditions, column B re-emulates Run~2 data using the \LtwoTauNNTag algorithm and scales the result to Run~3 conditions, and column C is the evaluated rate in Run~3. The instantaneous luminosities used were 1.68, 2.00, and $2.20 \times 10^{34} \percms$, respectively. The rates are inclusive calculations not excluding shared contributions from other algorithms or paths. The statistical uncertainties are negligible with respect to the significant digits reported. \label{tab:L2TauNNTag_comparison_tauhtauh_path}}
   \smallskip
   \begin{tabular}{c  c  c  c}
   Path & A [\kHzns] & B [\kHzns] & C [\kHzns] \\
   \hline
   Di-\tauh    & 6.1 & 5.4 & 5.5 \\
   Single-\tauh & 2.0 & 1.9 & 1.4 \\ 
   \end{tabular}
   \end{table}
      
\begin{figure*}[tp!]
   \centering
     {\includegraphics[width=0.45\textwidth]{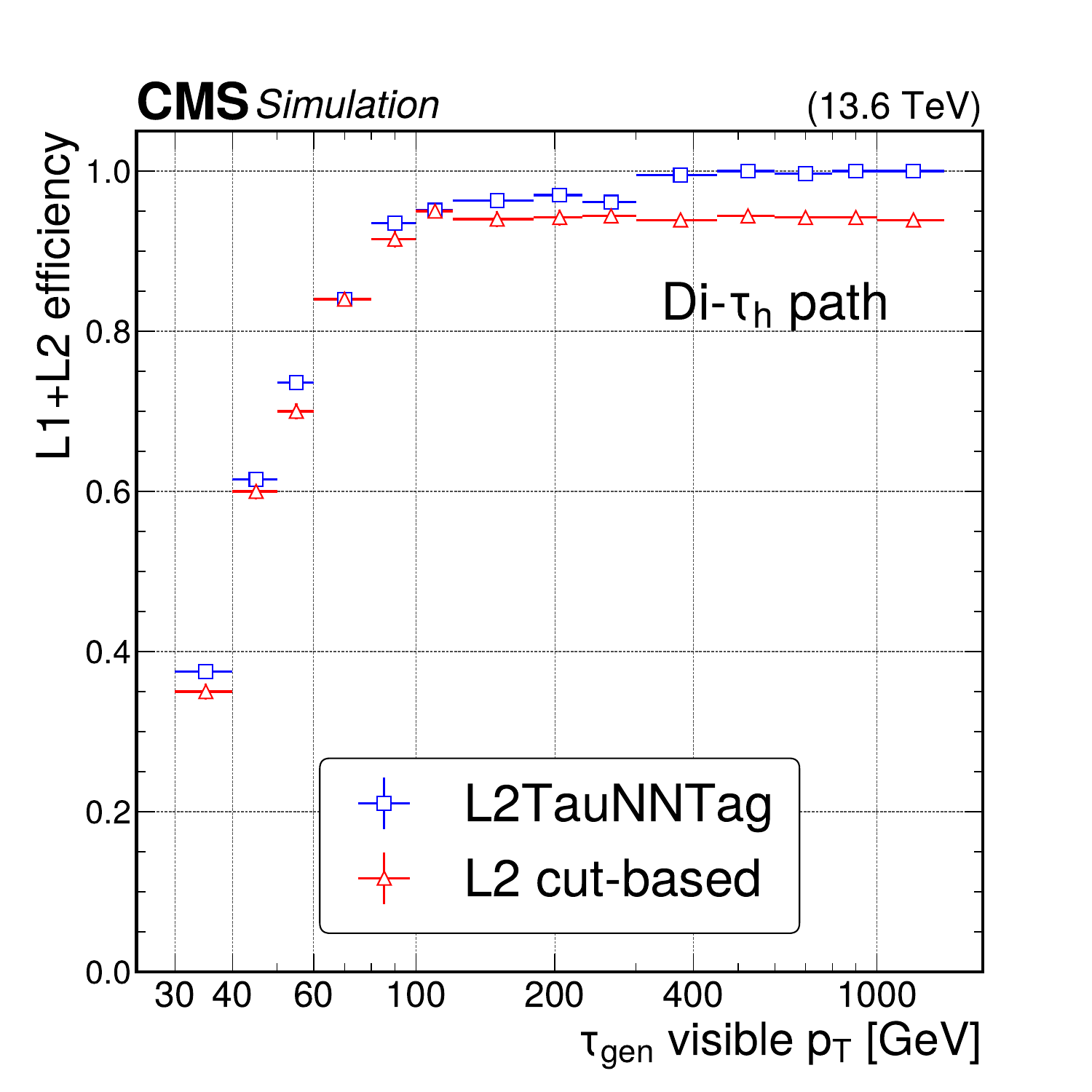}}
     {\includegraphics[width=0.45\textwidth]{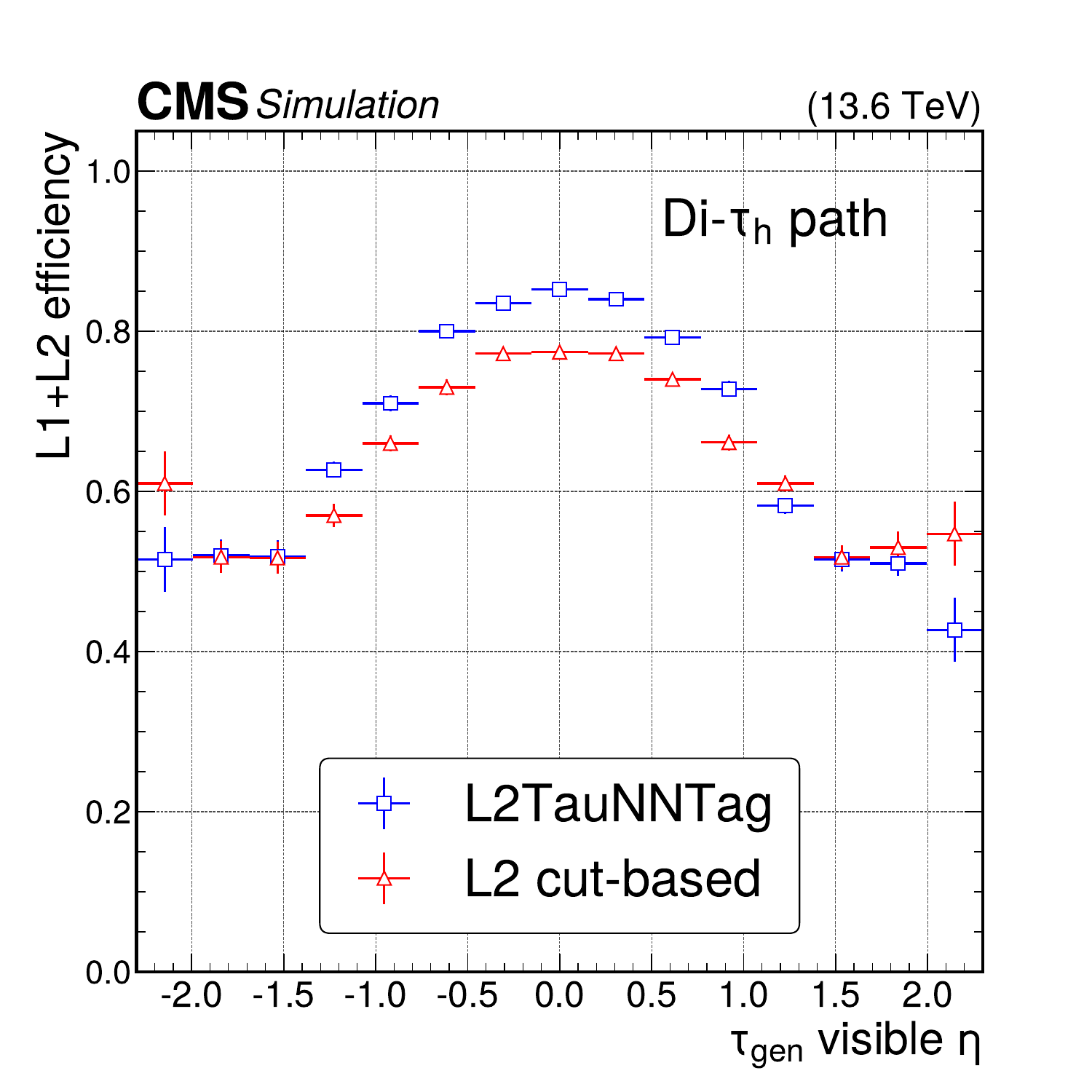}}
     \caption{Performance of the \LtwoTauNNTag in the di-\tauh HLT path, using simulated \PHtoPGtPGt and BSM \PZprtoPGtPGt events. The absolute efficiency of the reconstructed L2 \tauh candidates as a function of the visible generator-level \tauh \pt (left) and \penEta (right) are shown, where ``visible'' refers to the fact that the contribution of neutrinos is not taken into account. The uncertainties shown by the vertical bars are from the number of events available in the sample, while the horizontal bars show the bin width. Some of the vertical bars are smaller than the markers and are not shown.}\label{fig:L2TauNNTag_tauhtauh}
\end{figure*}

The efficiency of selecting genuine hadronic tau leptons using the di-\tauh HLT path is compared between the L2+L2.5 cut-based and \LtwoTauNNTag algorithms in Fig.~\ref{fig:L2TauNNTag_tauhtauh}, using simulated \PHtoPGtPGt and BSM \PZprtoPGtPGt events which were not used in the training.
The mixture of events from these two processes provides a sufficient sample size in a wide \pt range.
The absolute efficiency is calculated using the \pt-leading \tauh candidate and contains both the L1 and L2 steps, accounting for the trigger efficiency. 
In the left plot, the absolute efficiency of the \LtwoTauNNTag surpasses the L2+L2.5 cut-based approach in almost the full range of visible generator-level \tauh \pt, where ``visible'' refers to the fact that the contribution of neutrinos is not taken into account.
The efficiency jump at 300\GeV is due to a threshold selection, above which all candidates are accepted.
This originates from the fact that the L1 \tauh \pt value saturates above 250\GeV due to hardware constraints and is used as an input to the \LtwoTauNNTag.
Although the absolute efficiency of the L2+L2.5 cut-based approach is slightly higher than \LtwoTauNNTag in the high-\penEta region, an improvement in efficiency from the \LtwoTauNNTag can be seen in the central region of the visible generator-level \penEta, which most electroweak and targeted BSM physics processes favor.
The efficiency reduction in the high-\penEta region is due to the low number of events available in the training samples in that region.
Overall, the \LtwoTauNNTag results in better performance for the di-\tauh HLT path than the L2+L2.5 cut-based approach.

A comparison between the \LtwoTauNNTag and the L2 cut-based approach is also performed for the single-\tauh HLT path, using simulated \PHtoPGtPGt and BSM \PZprtoPGtPGt events that were not used in the training.
The introduction of the \LtwoTauNNTag step reduces the event processing time by roughly 40\%, while keeping the event rate unchanged with respect to the previous L2 cut-based approach.

\begin{figure*}[hbtp!]
   \centering
     {\includegraphics[width=0.45\textwidth]{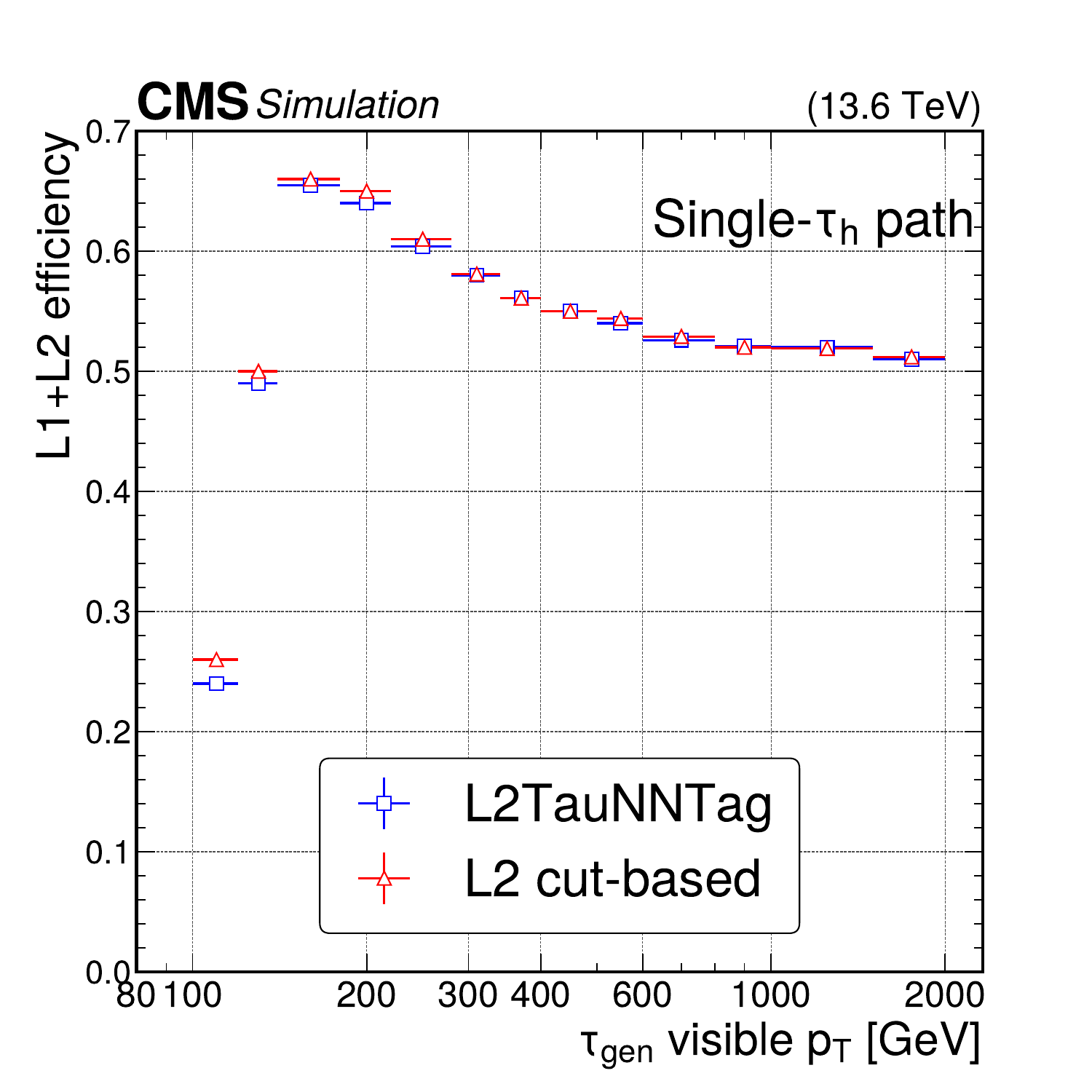}}
     {\includegraphics[width=0.45\textwidth]{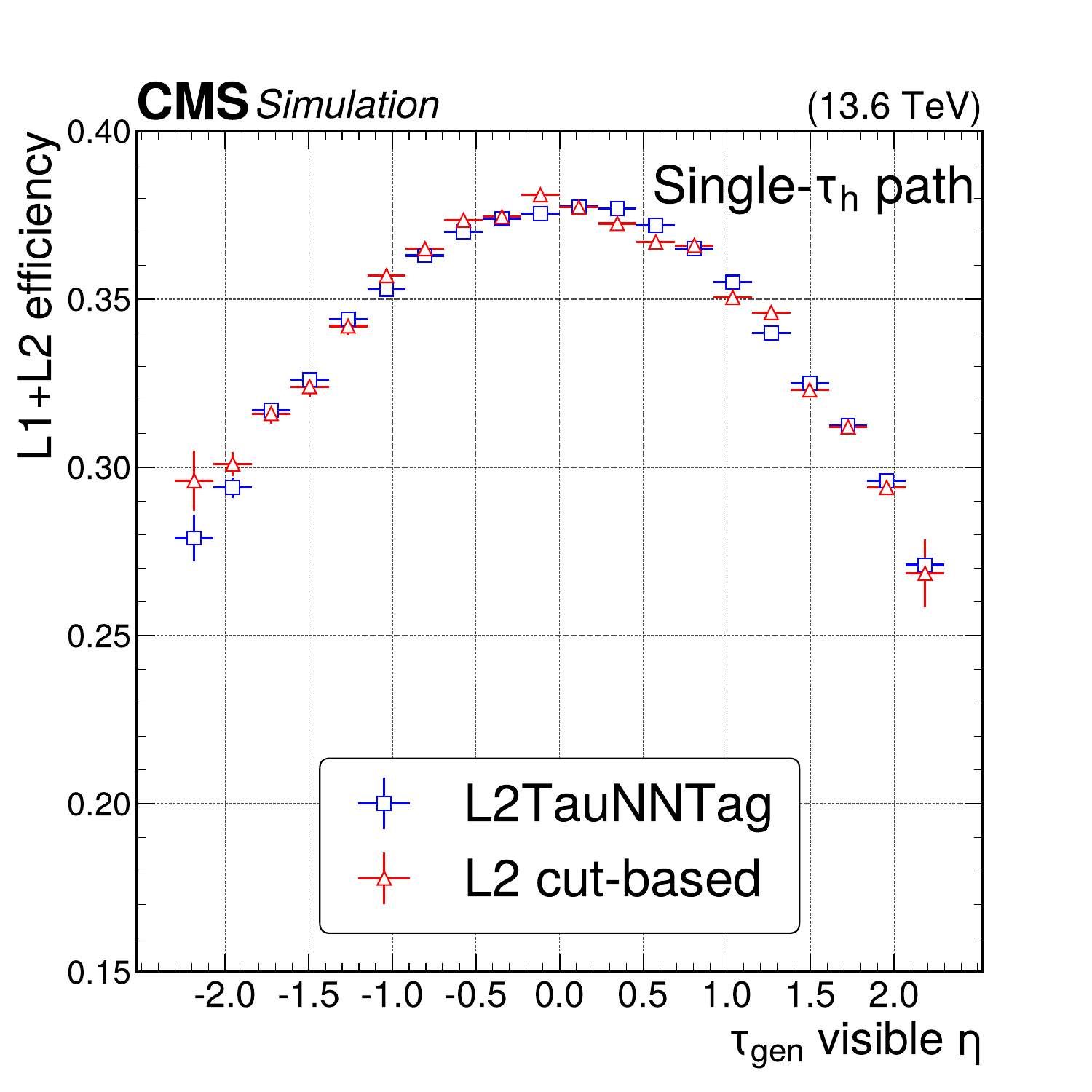}}
     \caption{Performance of the \LtwoTauNNTag in the single-\tauh HLT path, using simulated \PHtoPGtPGt and BSM \PZprtoPGtPGt events. The absolute efficiency of the reconstructed L2 \tauh candidates as a function of the visible generator-level \tauh \pt (left) and \penEta (right) are shown, where ``visible'' refers to the fact that the contribution of neutrinos is not taken into account. The uncertainties shown by the vertical bars are from the number of events available in the sample, while the horizontal bars show the bin width. Some of the vertical bars are smaller than the markers and are not shown.}\label{fig:L2TauNNTag_singletauh}
\end{figure*}

{\tolerance=1600
The efficiency of the single-\tauh HLT path is compared between the L2 cut-based and \LtwoTauNNTag algorithms in Fig.~\ref{fig:L2TauNNTag_singletauh}. The reduced efficiency in the high-\pt region of the left plot is caused by an L1-to-HLT \dR-matching requirement, which can fail at high \pt when multiple \tauh candidates are present but only one is reconstructed by either the L1 or HLT. Subsequently, L1-to-HLT matching is not required in the path itself, and the reduced efficiency is understood as an artifact of the selection.
Given the high efficiency already obtained in the L2 cut-based approach, the \LtwoTauNNTag was introduced in the single-\tauh path to reduce the processing time while keeping the efficiencies and rates at a similar level.
\par}

While both the di-\tauh and single-\tauh HLT paths have been updated to use the \LtwoTauNNTag, the paths have different requirements. In the di-\tauh HLT path, two \tauh candidates are required to pass a neural network discriminant value which was optimized for maximum efficiency improvement. In the single-\tauh HLT path, only one \tauh candidate is required, and it must pass a tighter discriminant value to retain the established efficiency. The differences in efficiency between these two HLT paths are due to these differing object requirements, as displayed in Figs.~\ref{fig:L2TauNNTag_tauhtauh}~and~\ref{fig:L2TauNNTag_singletauh}.

\section{\texorpdfstring{\DeepTau implementation in the HLT}{DeepTau implementation in the HLT}}\label{sec:deeptau}

The \DeepTau algorithm is a multiclass \tauh candidate identification algorithm based on a convolutional deep neural network, originally designed for use on offline objects~\cite{CMS:2025kgf}.
The primary goal of \DeepTau is to effectively discriminate hadronic tau lepton decays from all three main backgrounds: quark or gluon jets, electrons, and muons. To accomplish this, it is used on PF jets after they have undergone HPS reconstruction~\cite{CMS:2022prd}.
The online HLT version of \DeepTau adopts the same approach, except that only the discriminant against jets is used.
\DeepTau combines information from the high-level features of the reconstructed \tauh candidate with the low-level information from the inner tracker, calorimeters, and muon subdetectors using PF candidates, electrons, and muons reconstructed within the \tauh candidate's isolation cone.

To ensure any analysis with final-state hadronic tau lepton decays could benefit from the \DeepTau algorithm, the training is performed on a balanced mixture of electron, muon, \tauh candidates, and jets from MC datasets. Simulated samples of DY, \ttbar, and \PWj are used for genuine hadronic tau lepton decays. Simulated \ttbar decays with one or two light leptons (\Pe or \PGm) and QCD samples are used for electrons, muons, and jets misidentified as hadronic tau lepton decays, respectively.

In order to train the \DeepTau algorithm such that it is suitable for a wide kinematic range, balance in the type of the \tauh candidates and their respective \pt and \penEta is crucial. Available samples show diverse compositions of \tauh candidate decay modes and \PtDashEta distributions, with more events available in the low-\pt and central-\penEta regions, and fewer events in the high-\pt and high-\penEta regions. In order to ensure a uniform contribution from different samples, decay modes, and \PtDashEta bins, additional weights are used in the training.

\subsection{Network architecture}
The HLT \DeepTau algorithm largely follows the offline version.
The inputs are formatted in the \EtaDashPhi plane in the same way as the offline \DeepTau algorithm, with an ``isolation cone'' composed of an outer grid of $21{\times}21$ cells with \dEtaTimesdPhi grid sizes of $0.05{\times}0.05$, and a ``signal cone'' with dense inner grid of $11{\times}11$ cells with \dEtaTimesdPhi grid sizes of $0.02{\times}0.02$.
To reduce the computing cost for the algorithm, the two grids are not allowed to overlap.
Using a denser inner grid exploits the signal tau lepton decay products that are usually within a $\dR \approx 0.1$ cone on the \EtaDashPhi plane. A dense inner grid also copes well with high-\pt jets whose constituents are collimated.
The radius of the signal cone is defined as $\Rsig = 3\GeV/\pt$, with minimum and maximum allowed values of 0.05 and 0.1, while the isolation cone is defined as the remaining range from the edge of \Rsig to $R = 0.5$.
Three different subnetworks are constructed to be able to independently process the inputs from the high-level variables, the outer cells, and the inner cells.
The outputs from the three subnetworks are concatenated and passed through four fully connected layers with 200 nodes each. The results are collected in a final layer with four nodes that yield the outputs.
A softmax activation function~\cite{GoodBengCour16} is then applied to yield estimates of the probabilities for the \tauh candidate to come from each of the four target classes: \tauh, jet, \Pe, and \Pgm.

For usage of \DeepTau at the HLT, some simplifications in the inputs of the network were made.
Because an individual HLT path must rely on the objects it produces, it is not possible to access reconstructed electrons and muons for every \tauh candidate at the HLT. Consequently, the electron and muon discriminants are not used in the online \DeepTau algorithm. Leptons being misidentified as \tauh candidates are however not a leading background.
Similarly, association and quality fits of the primary vertex for online PF candidates are not performed due to the online processing time constraint. Additionally, several track-related variables, such as \dz, are not used in the online \DeepTau algorithm either. Despite these limitations, the \DeepTau network still performs successfully, retaining high performance with the features available online.

\subsection{\texorpdfstring{Performance of the \DeepTau algorithm}{Performance of the DeepTau algorithm}}
The performance of several \tauh HLT paths equipped with online \DeepTau identification is presented in terms of event rate and HLT path efficiency.
The working point is optimized for each HLT path to maximize the signal efficiency under the constraint that the HLT path retains an event rate similar to its Run~2 value.
The optimization of the \DeepTau identification thresholds is performed as a function of the \tauh candidate \pt.

\begin{figure*}[hbtp!]
   \centering
     {\includegraphics[width=0.45\textwidth]{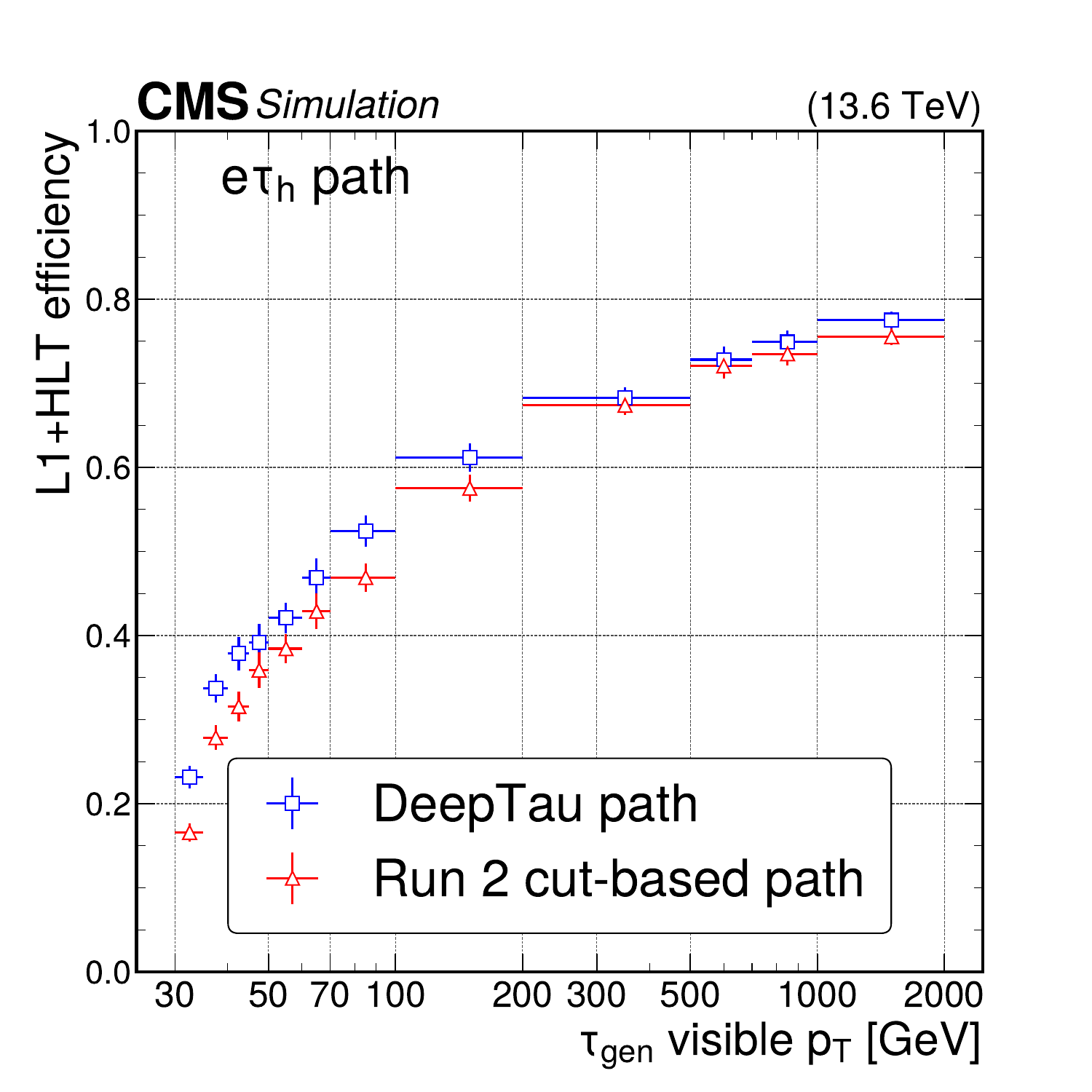}}
     {\includegraphics[width=0.45\textwidth]{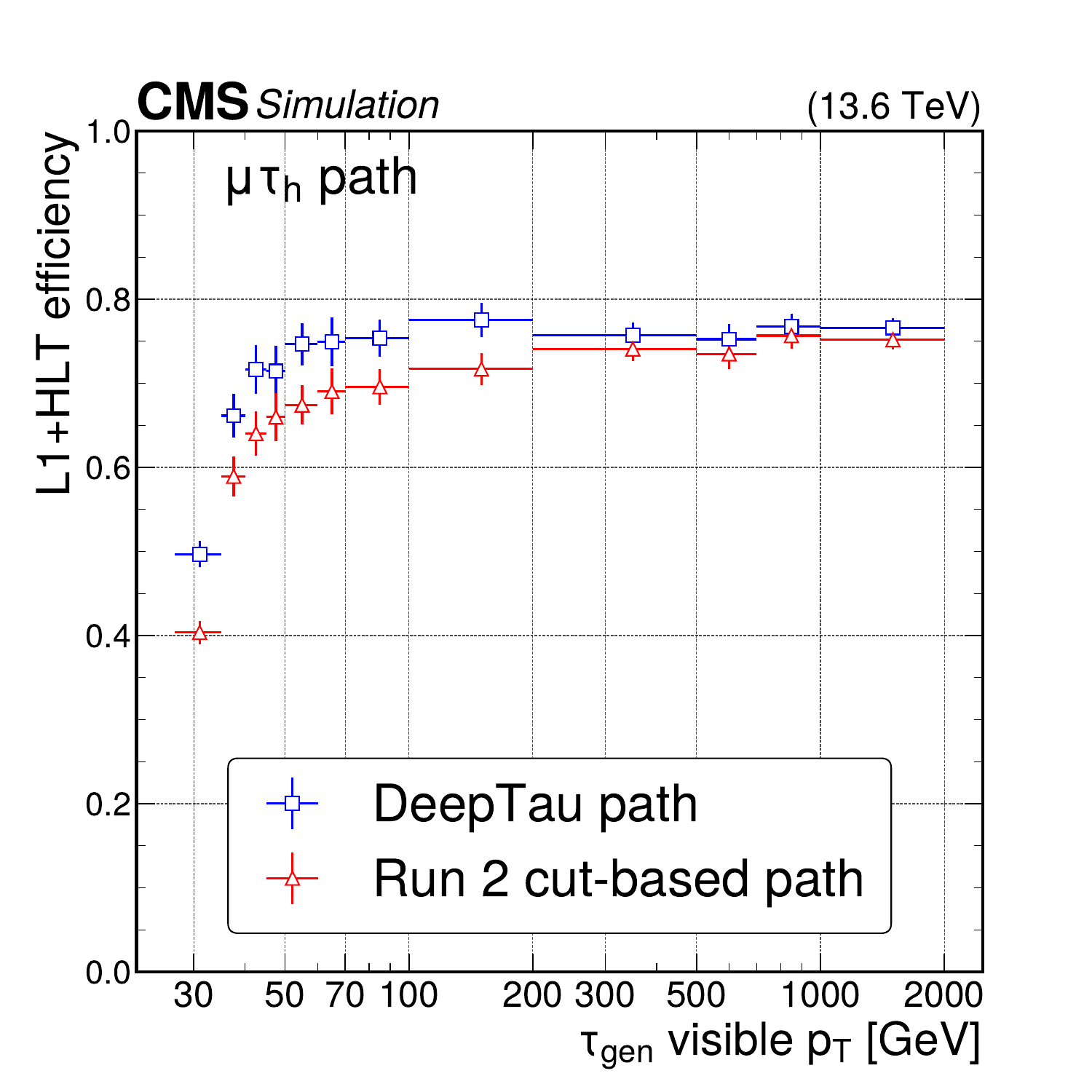}}
     {\includegraphics[width=0.45\textwidth]{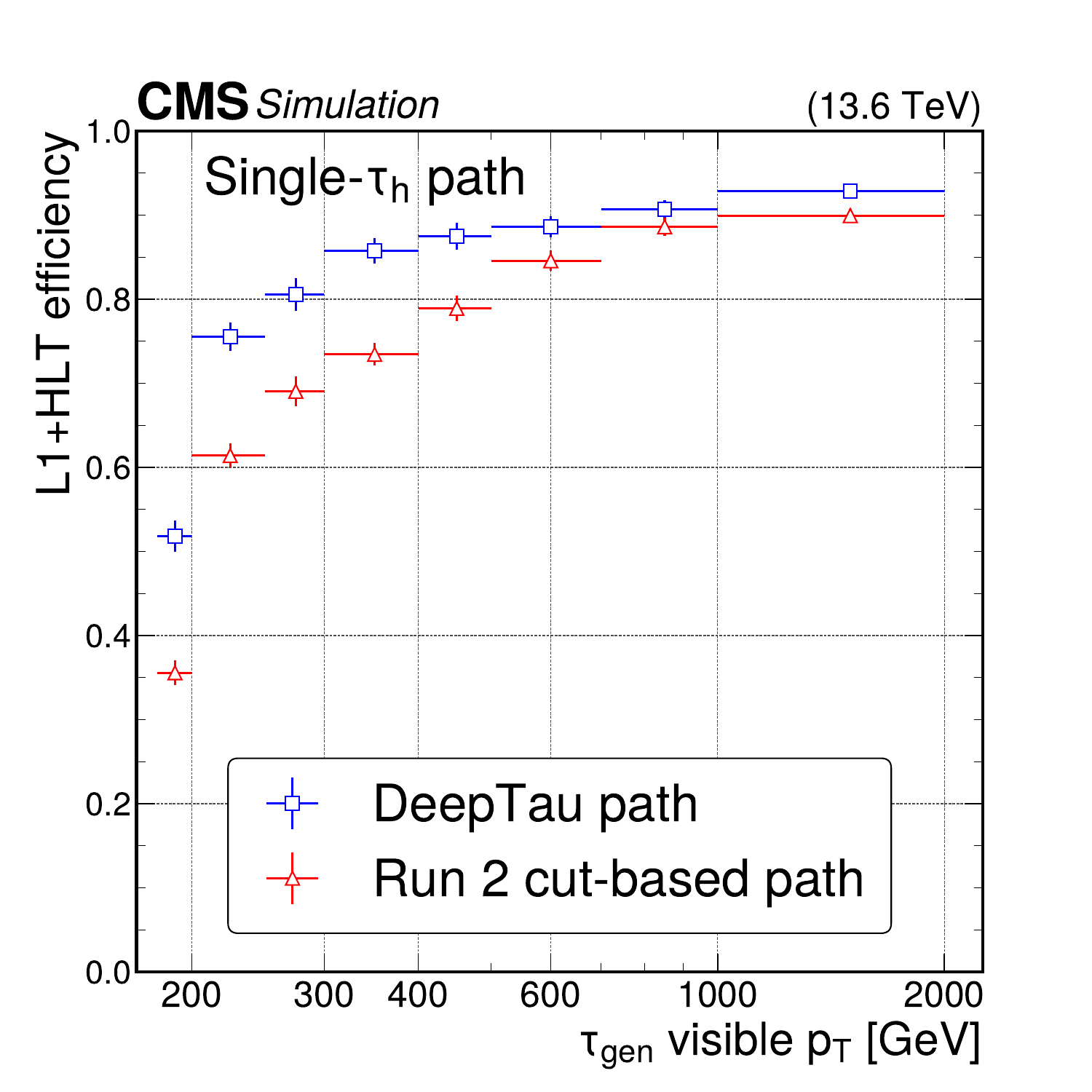}}
     {\includegraphics[width=0.45\textwidth]{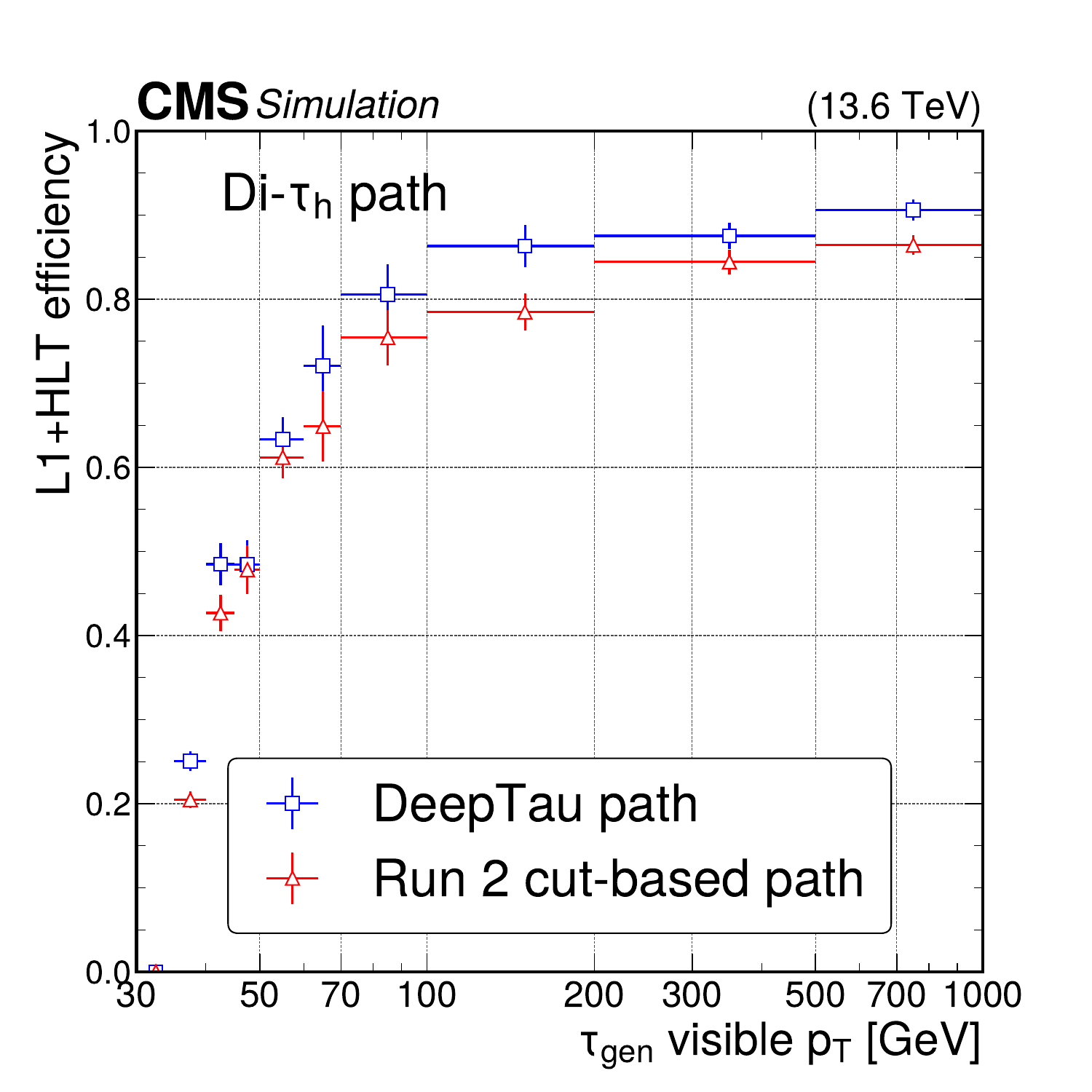}}
     \caption{Total L1\textplus HLT path efficiency of the \etauh (upper left), \mutauh (upper right), single-\tauh (lower left), and di-\tauh (lower right) HLT paths as a function of the visible generator-level \tauh \pt, where ``visible'' refers to the fact that the contribution of neutrinos is not taken into account. The \PHtoPGtPGt and BSM \PZprtoPGtPGt samples are used in the evaluation. The uncertainties shown by the vertical bars are from the number of events available in the sample, while the horizontal bars show the bin width. Some of the vertical bars are smaller than the markers and are not shown.}\label{fig:eff_vis_pttau}
\end{figure*}

Several \tauh HLT path rates are shown in Table~\ref{tab:DeepTau_rate}. 
The estimated \DeepTau rates are computed using a dataset dedicated to rate estimates, collected in October 2018, corresponding to 2\unit{hours} of data taking at an average pileup of 47. This rate was then scaled to the expected Run~3 conditions, targeting a pileup of 60. The observed \DeepTau rates are computed using a dataset collected in November 2022, corresponding to 4.5\unit{hours} of data taking at an average pileup of 54.
A comparison of the overall efficiency between \DeepTau and the Run~2 cut-based approach is shown in Fig.~\ref{fig:eff_vis_pttau},
where the \LtwoTauNNTag filter is disabled in order to focus on the \DeepTau performance.
The efficiencies are evaluated using MC samples of \PHtoPGtPGt and BSM \PZprtoPGtPGt events, simulated with early Run~3 conditions.
In the presence of two \tauh candidates, the one with greater \pt is used in the calculation.
The HLT paths using \DeepTau are shown to outperform those using the Run~2 cut-based selection in the majority of the \pt range, while maintaining a similar event rate.

\begin{table}[htbp]
  \centering
	\topcaption{Rate estimation and observation for several \tauh HLT paths. Similarly to Table~\ref{tab:L2TauNNTag_comparison_tauhtauh_path}, column A scales Run~2 data collected by the cut-based algorithm to Run~3 conditions, column B re-emulates Run~2 data using the \LtwoTauNNTag algorithm and scales the result to Run~3 conditions, and column C is the evaluated rate in Run~3. The instantaneous luminosities used for the Run~2 estimation, Run~3 projection, and Run~3 evaluation were 1.68, 2.00, and $2.20 \times 10^{34} \percms$, respectively. The rates are inclusive calculations not excluding shared contributions from other algorithms or paths. The statistical uncertainties are negligible with respect to the significant digits reported.\label{tab:DeepTau_rate}}
    \smallskip
  \begin{tabular}{c  c  c  c}
	HLT path & A [\Hzns] & B [\Hzns] & C [\Hzns] \\
        \hline
	\etauh & 10 & 9 & 10\\
	\mutauh & 5 & 5 & 5\\
	Single-\tauh & 14 & 13 & 15\\
	Di-\tauh & 55 & 46 & 51\\
\end{tabular}
\end{table}
   
\section{\texorpdfstring{Overall performance of the \tauh HLT paths}{Overall performance of the hadronic tau HLT paths}}\label{sec:performance}
The performance studies presented in this Section use \PpPp collision data from 2022--2023. For all HLT paths, a ``tag-and-probe'' technique is used to measure the trigger efficiency~\cite{CMS:2018jrd,CMS:2019zly}. With this method, one first selects a tag object that satisfies stringent requirements, and then selects the probe object, whose efficiency is measured. In the case of \tauh HLT paths, the tag object used is always a muon, which is a well-isolated object highly distinguishable against jets. To perform this measurement in situ, so-called monitoring HLT paths are made which require one muon and one \tauh candidate. Each \tauh HLT path has a corresponding monitoring HLT path with the exact same \tauh candidate requirements. In this manner, the tag-and-probe method is applied to the monitoring HLT paths, and the measured efficiencies correspond to the signal HLT paths. This approach was validated using MC samples by comparing the signal HLT path efficiencies to the monitoring HLT path efficiencies collected with the tag-and-probe method, and the results were found to agree within uncertainties. 

\subsection{Efficiencies in experiment}
The efficiency is measured from a region enriched in DY events, where the \PZ boson decays into two \PGt leptons that subsequently decay into a \mutauh final state. The tag muon is required to have $\pt > 24\GeV$, $\abseta < 2.4$, and be isolated such that nearby calorimeter energy deposits sum to less than 15\% of the muon's \pt. Additionally, the tag muon is required to be within $\dR < 0.5$ of the muon reconstructed by the single muon HLT path, and pass the Medium ID for offline muons~\cite{CMS:2018rym}. The probe \tauh candidate has the offline selection of $\pt > 20\GeV$, $\abseta < 2.1$, and passes the following offline \DeepTau working points: VVLoose (Very Very Loose) against electrons, Tight against muons, and Medium against jets~\cite{CMS:2022prd}, summarized as ``Offline Medium WP DeepTau ID applied'' in relevant figures. Each working point corresponds to a specific trade-off between \tauh selection efficiency and background rejection rate. The muon and \tauh candidate must have opposite charges. A visible invariant mass window is required with $40 < m_{\text{vis}} (\PGm, \tauh) < 80\GeV$ to reduce the efficiency impact of energy loss from neutrinos. To reduce the background contamination from \PWj processes, the event is also required to have a transverse mass less than 30\GeV, which is defined by the muon and \ptvecmiss as $\mT(\vec{p}_\text{T}^\PGm,\ptvecmiss)=\sqrt{\smash[b]{2p_\text{T}^\PGm\ptmiss[ 1 - \cos\dPhi(\vec{p}_\text{T}^\PGm,\ptvecmiss) ]}}$. The denominator of the efficiency includes all the \mutauh events passing the above selections, and the numerator includes the events also passing the relevant monitoring HLT path with the offline \tauh candidate matched to the \tauh candidate reconstructed at the HLT with $\dR < 0.5$. The efficiency of the single-\tauh HLT path cannot be measured with this method due to its high \pt threshold, resulting in few suitable events. It is measured with an alternative strategy selecting events with large \ptmiss that include $\PW\to\tauh\PGnGt$ decays.

\begin{figure*}[hbtp!]
   \centering
     {\includegraphics[width=0.45\textwidth]{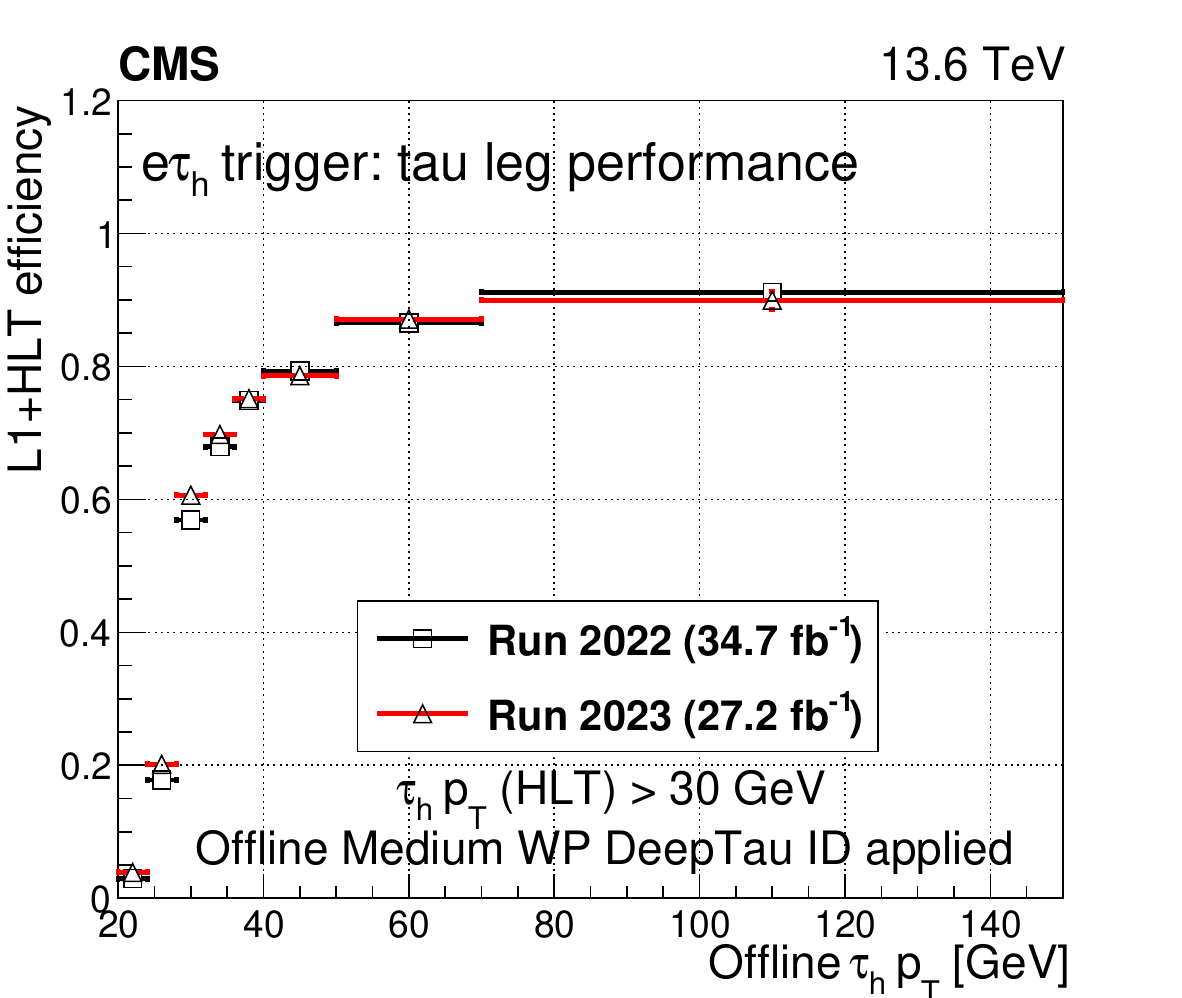}}
     {\includegraphics[width=0.45\textwidth]{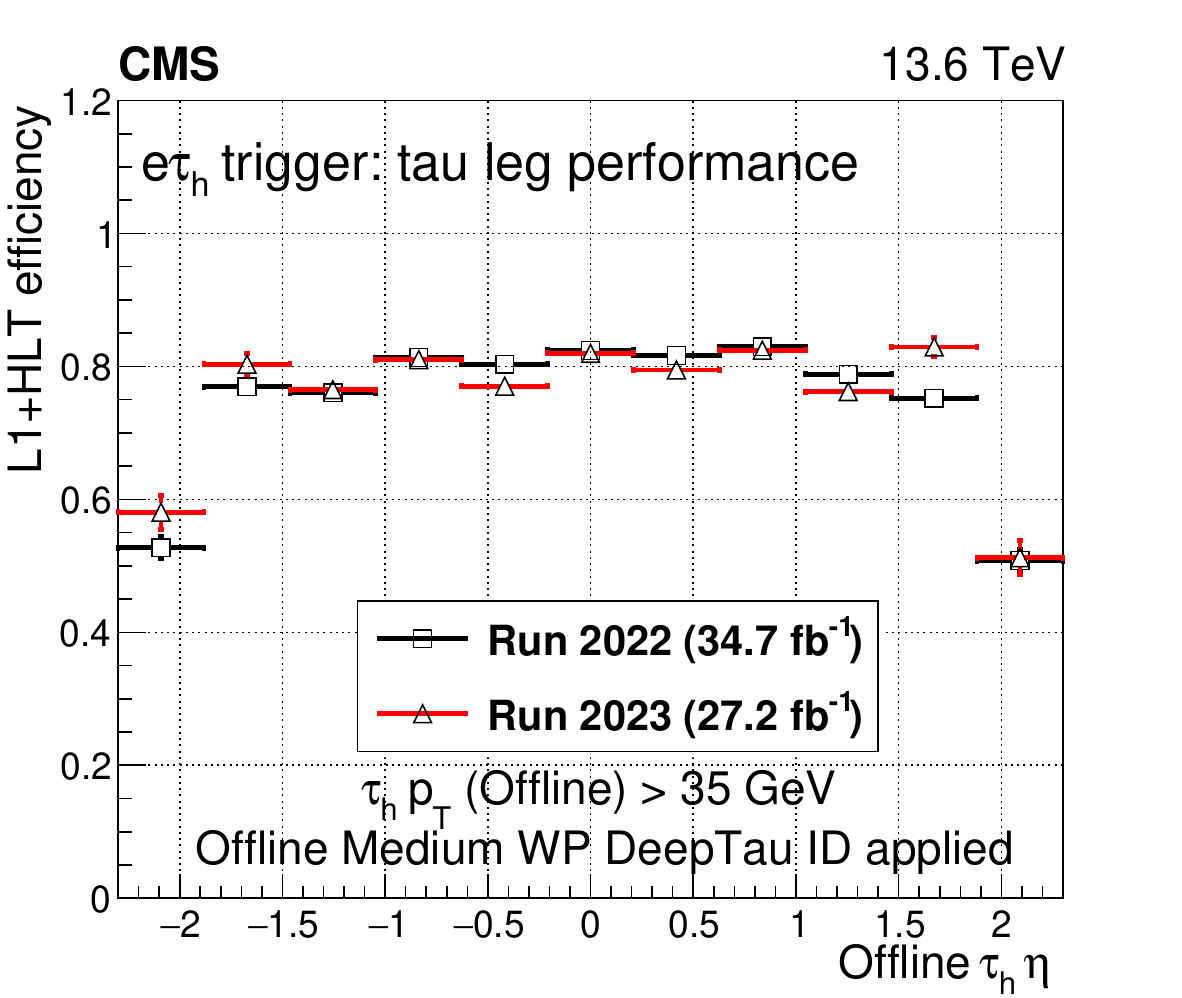}}
     {\includegraphics[width=0.45\textwidth]{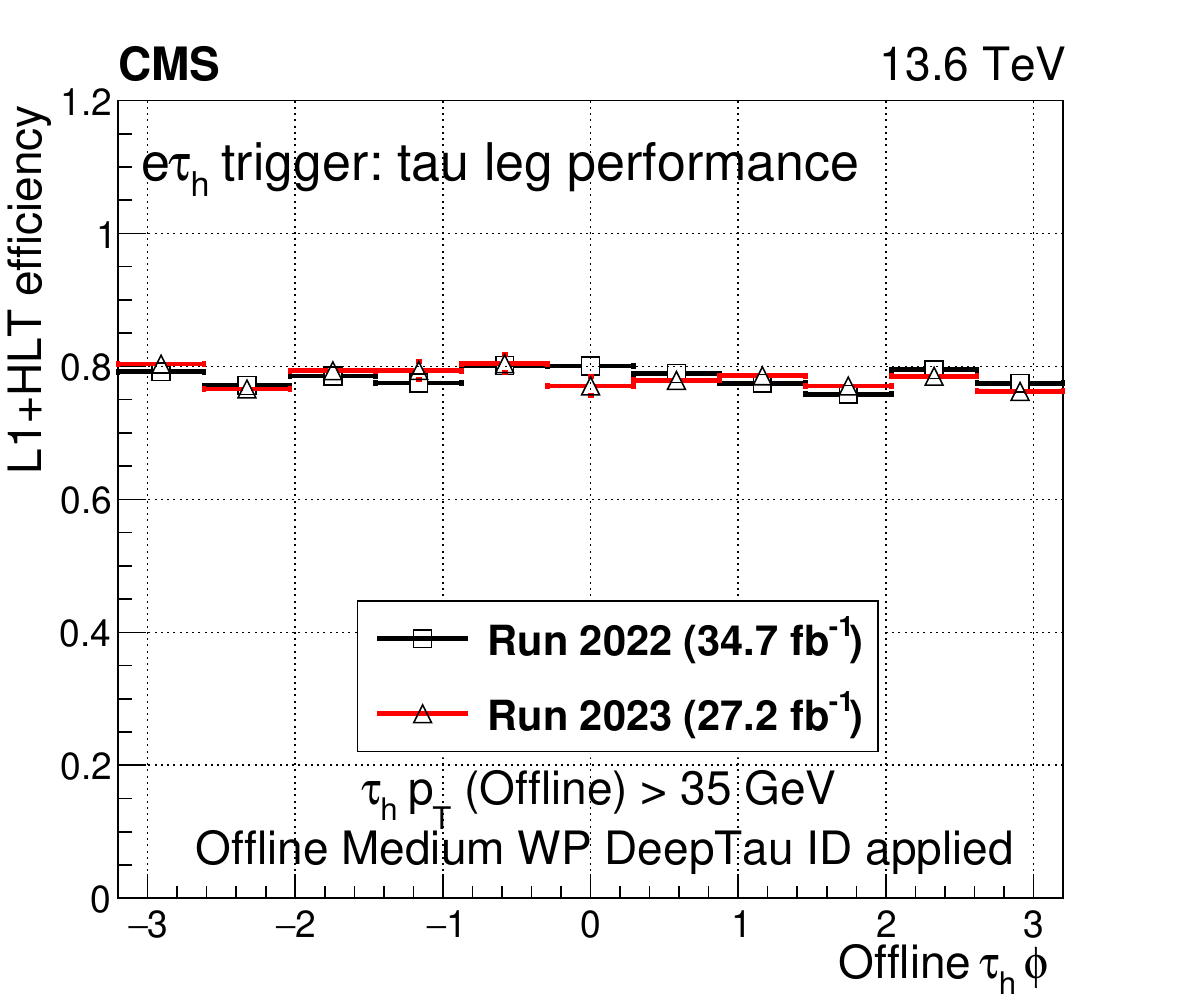}}
     {\includegraphics[width=0.45\textwidth]{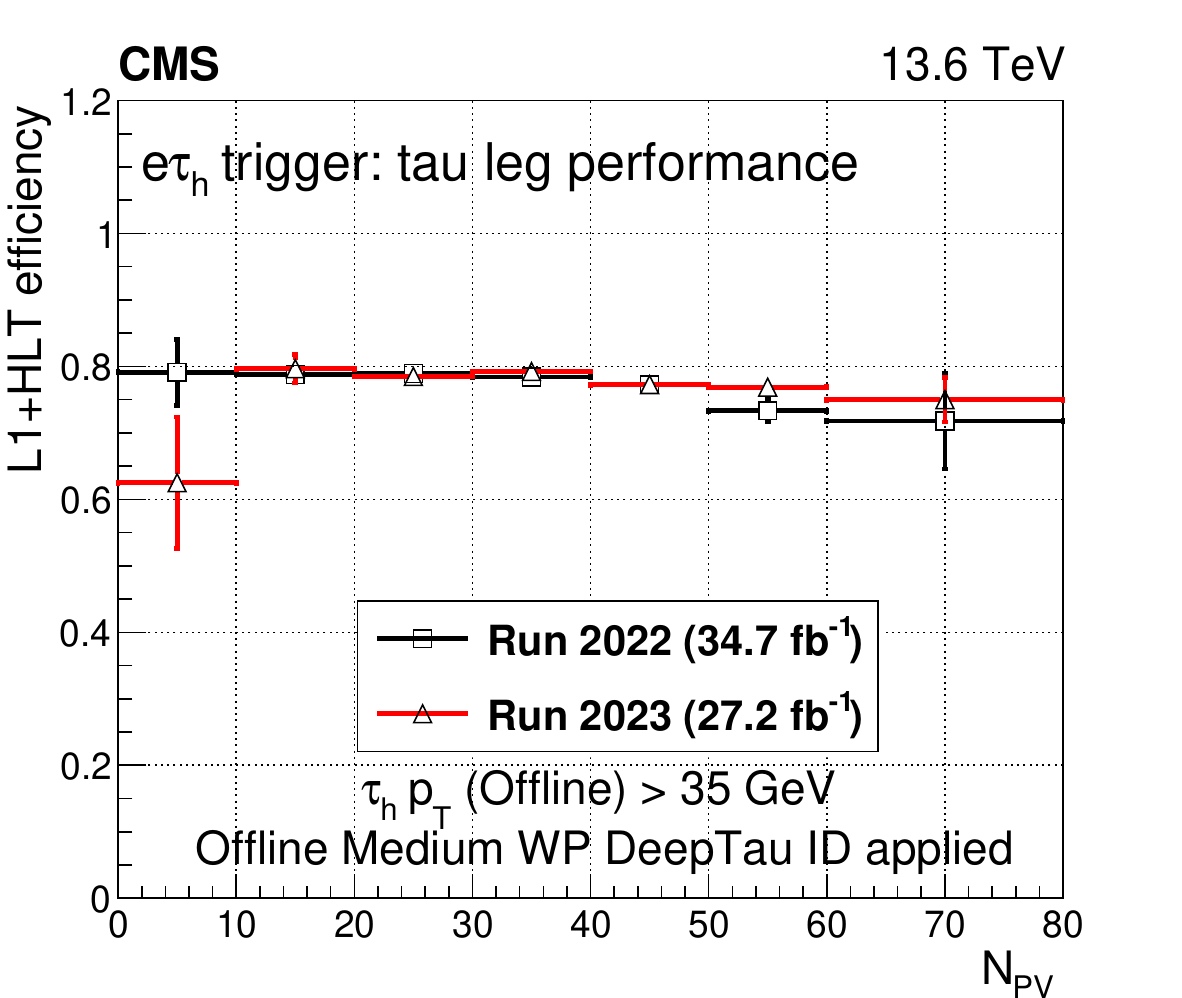}}
     \caption{A comparison of the L1\textplus HLT efficiency of the \etauh monitoring HLT path in 2022 and 2023 as a function of offline \tauh candidate \pt (upper left), \penEta (upper right), and \penPhi (lower left). The dependence on the \NPV is also shown (lower right). The uncertainties shown by the vertical bars are from the number of events available in the sample, while the horizontal bars show the bin width. Some of the vertical bars are smaller than the markers and are not shown.}\label{fig:etau_2022_2023}
\end{figure*}

\begin{figure*}[hbtp!]
   \centering
     {\includegraphics[width=0.45\textwidth]{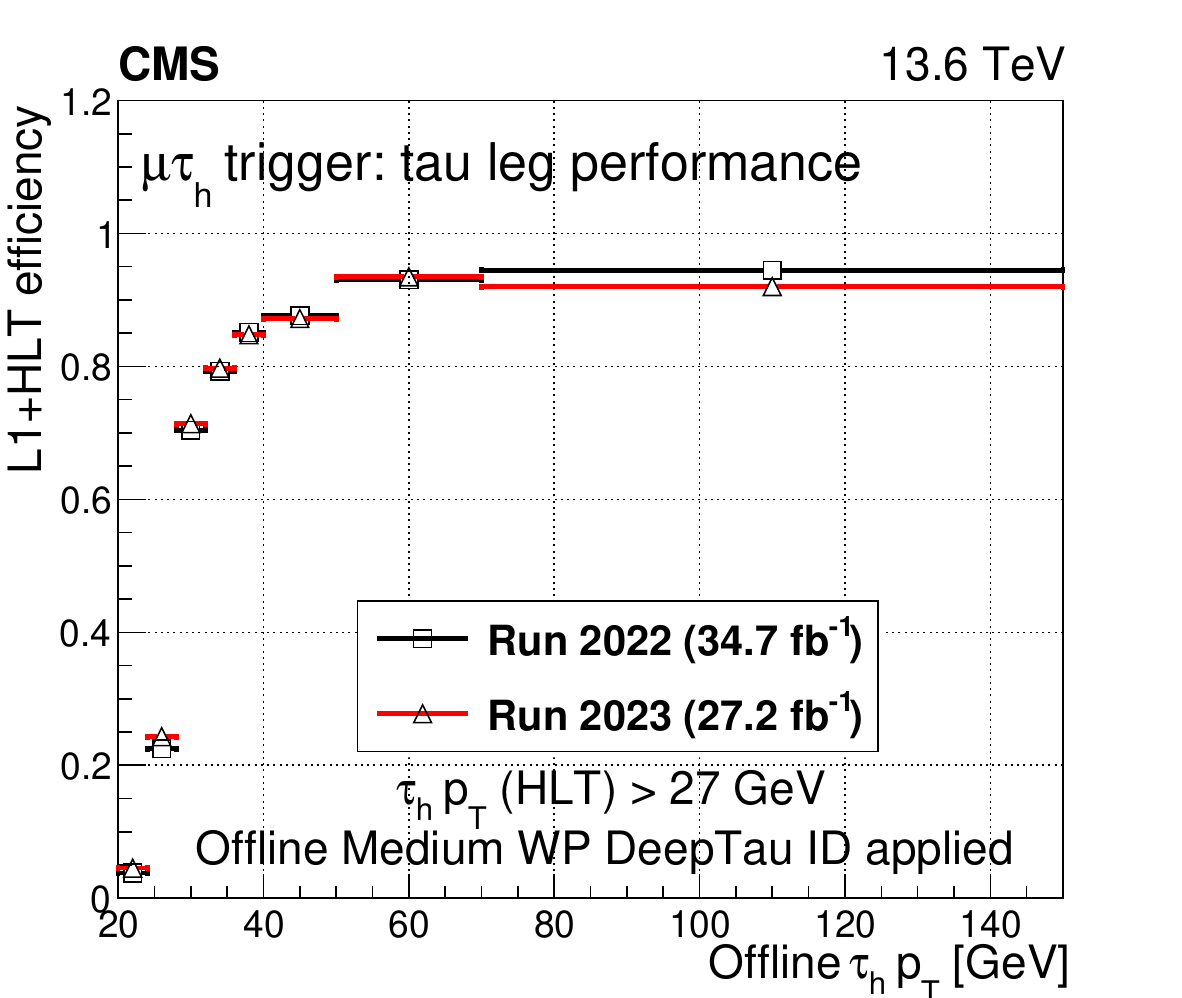}}
     {\includegraphics[width=0.45\textwidth]{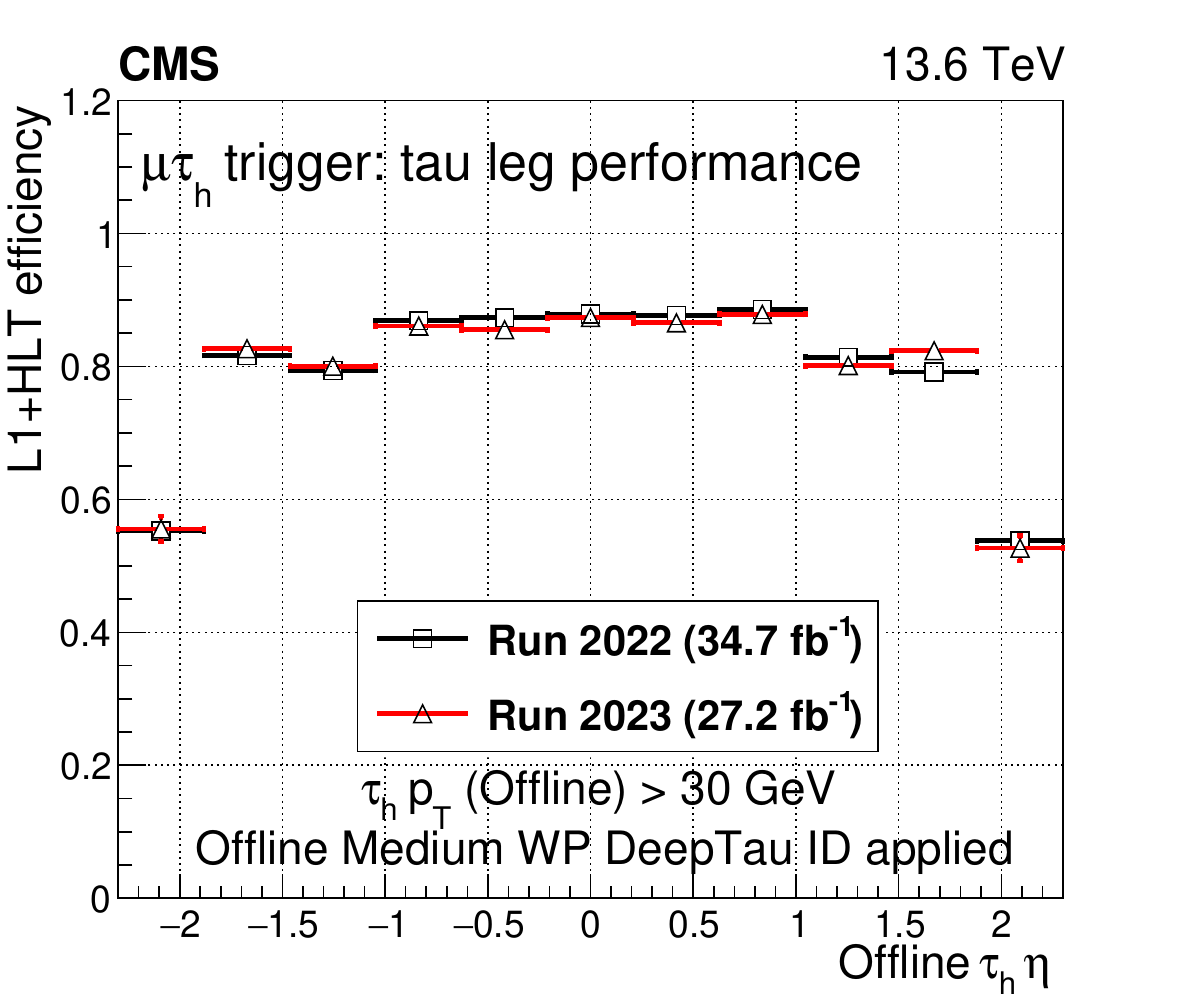}}
     {\includegraphics[width=0.45\textwidth]{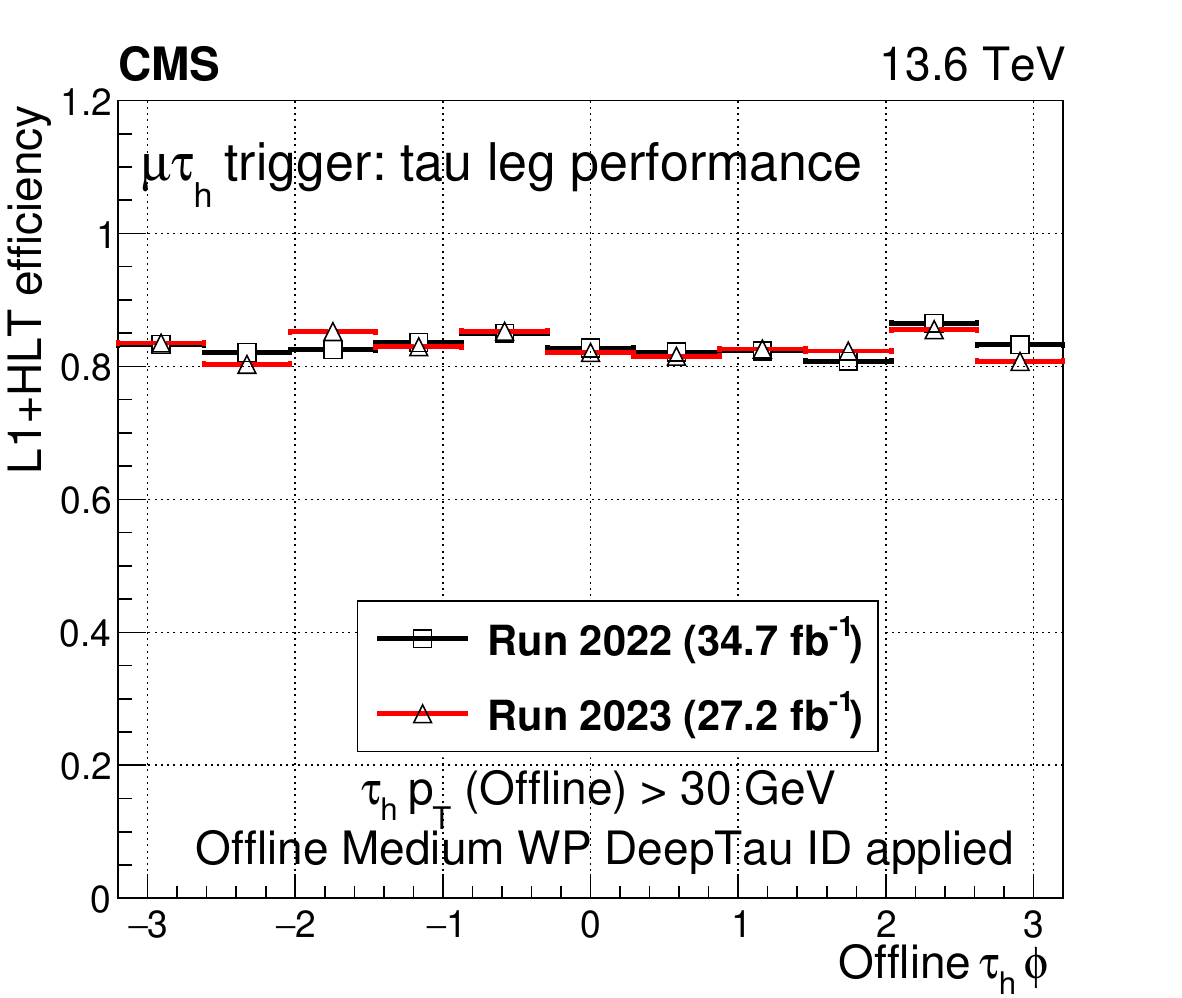}}
     {\includegraphics[width=0.45\textwidth]{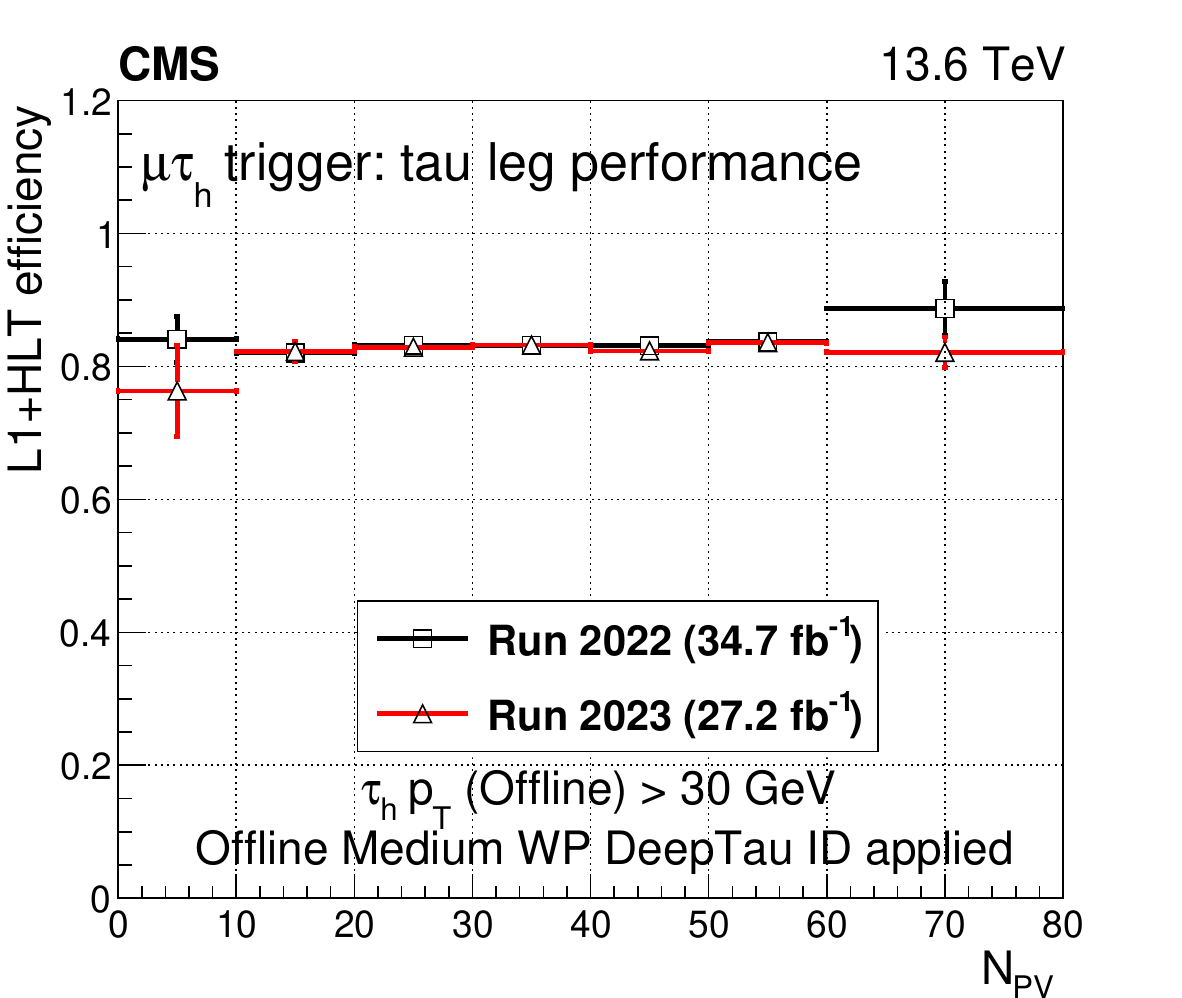}}
     \caption{A comparison of the L1\textplus HLT efficiency of the \mutauh HLT path in 2022 and 2023 as a function of offline \tauh candidate \pt (upper left), \penEta (upper right), and \penPhi (lower left). The dependence on the \NPV is also shown (lower right). The uncertainties shown by the vertical bars are from the number of events available in the sample, while the horizontal bars show the bin width. Some of the vertical bars are smaller than the markers and are not shown.}\label{fig:mutau_2022_2023}
\end{figure*}

The efficiencies of the \tauh leg in the \etauh and \mutauh HLT paths are shown in Figs.~\ref{fig:etau_2022_2023}~and~\ref{fig:mutau_2022_2023}, respectively, for 2022 and 2023.
The absolute path efficiency is calculated, which includes the L1+HLT \tauh candidate efficiency.
The \etauh (\mutauh) cross trigger has an online \tauh candidate \pt threshold of $30~(27)\GeV$; in addition to the offline selections for the \tauh candidate described above, the \tauh candidate is required to satisfy $\pt > 35~(30)\GeV$ for the efficiency measurements as a function of \penEta, \penPhi, and the number of primary vertices (\NPV). This reduces the impact of the low efficiency ``turn-on'' region of the HLT path, where events start to meet the threshold of passing the HLT path. 
The efficiency is robust against pileup, indicated by a stable \NPV plot, and the performance between the two years is comparable despite slightly different detector conditions.

\clearpage 

\clearpage 

\begin{figure*}[hbtp!]
   \centering
     {\includegraphics[width=0.45\textwidth]{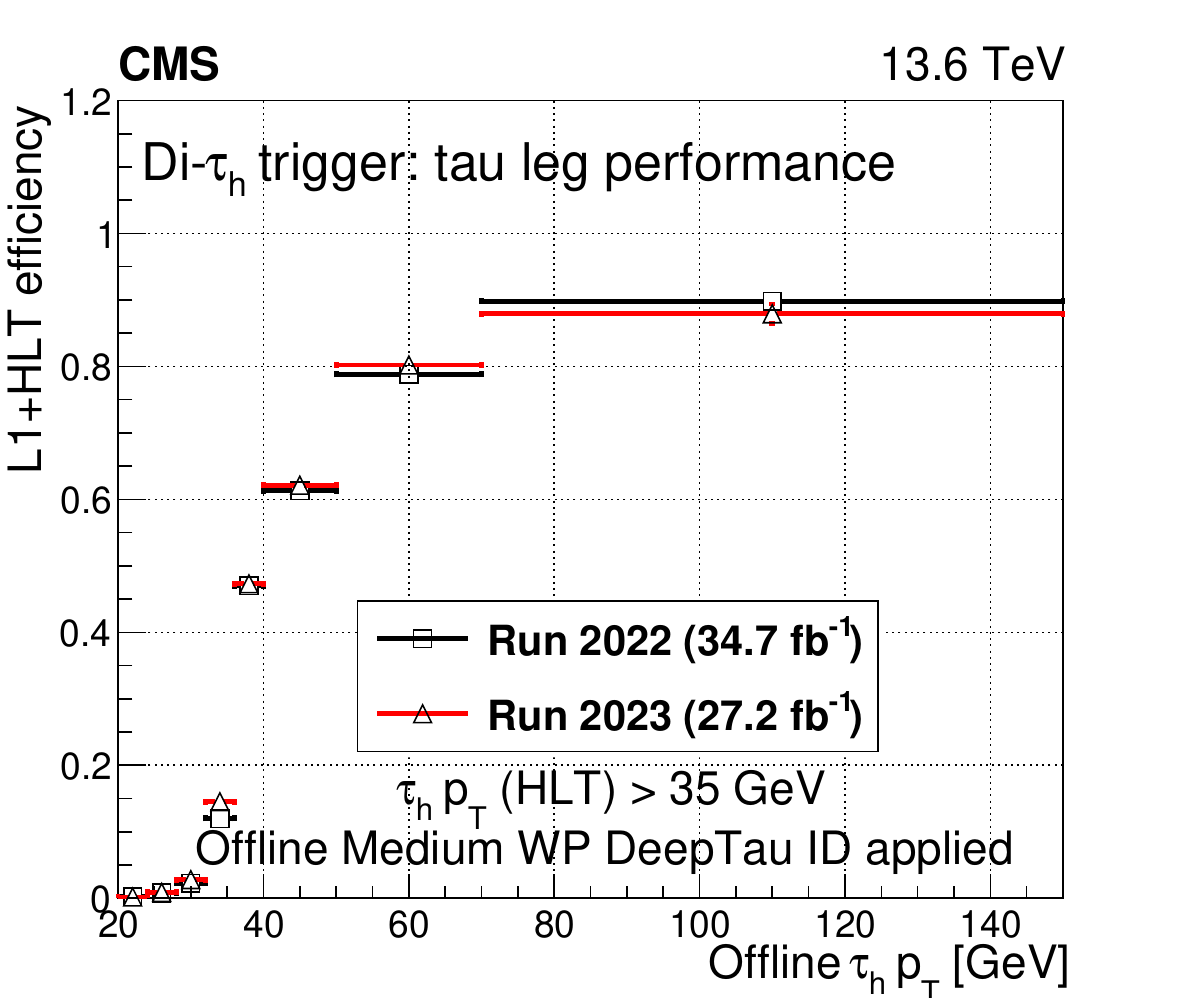}}
     {\includegraphics[width=0.45\textwidth]{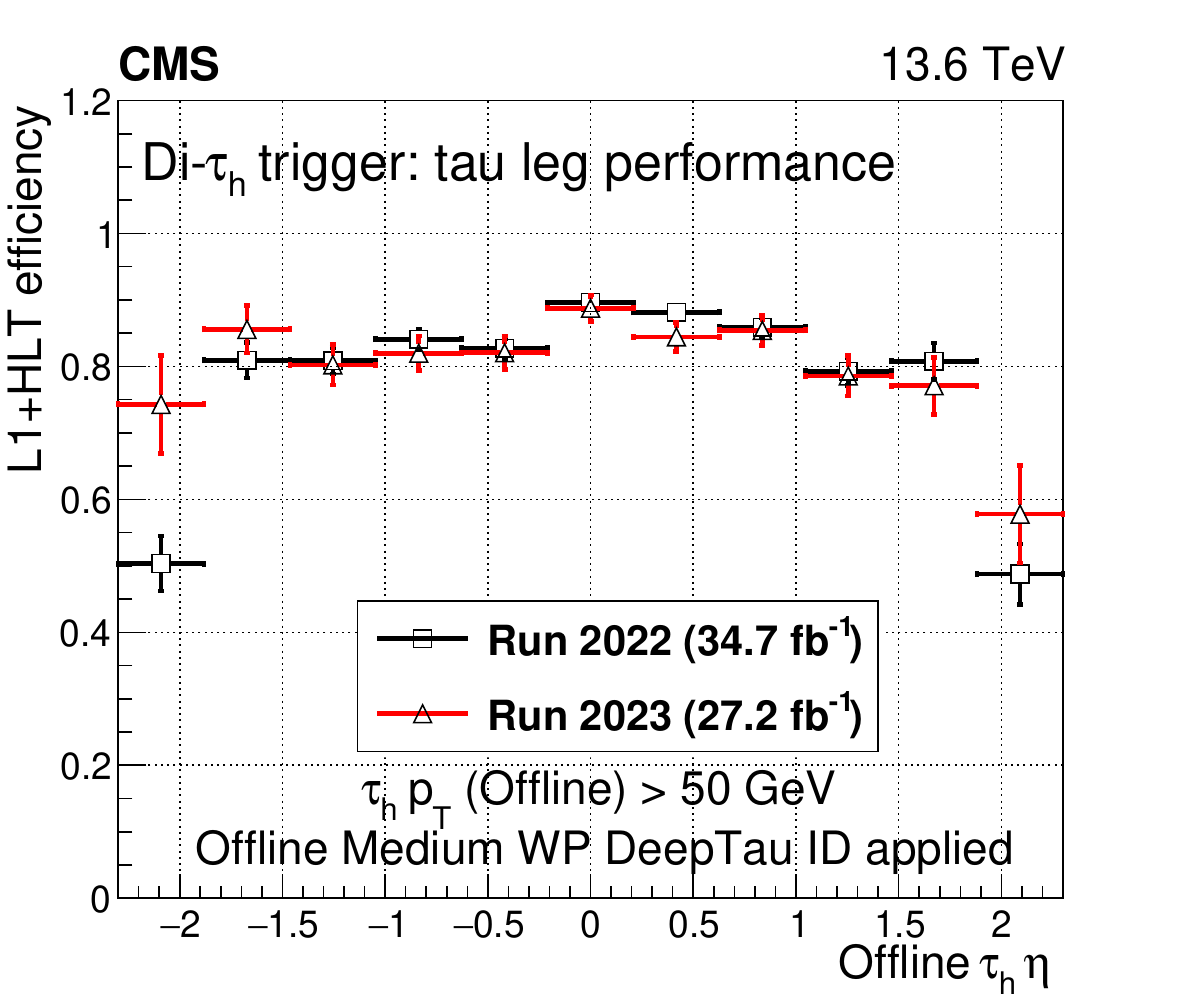}}
     {\includegraphics[width=0.45\textwidth]{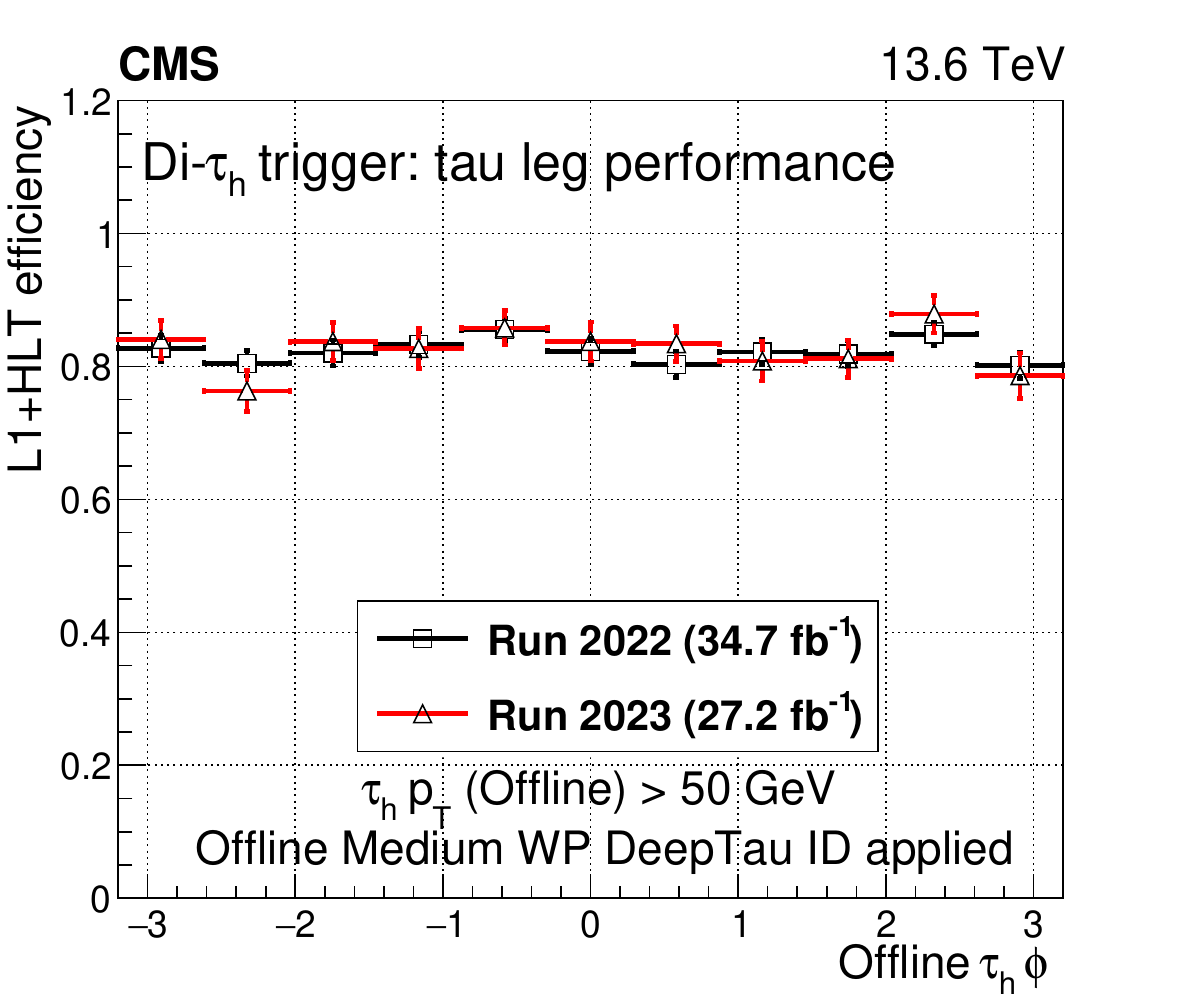}}
     {\includegraphics[width=0.45\textwidth]{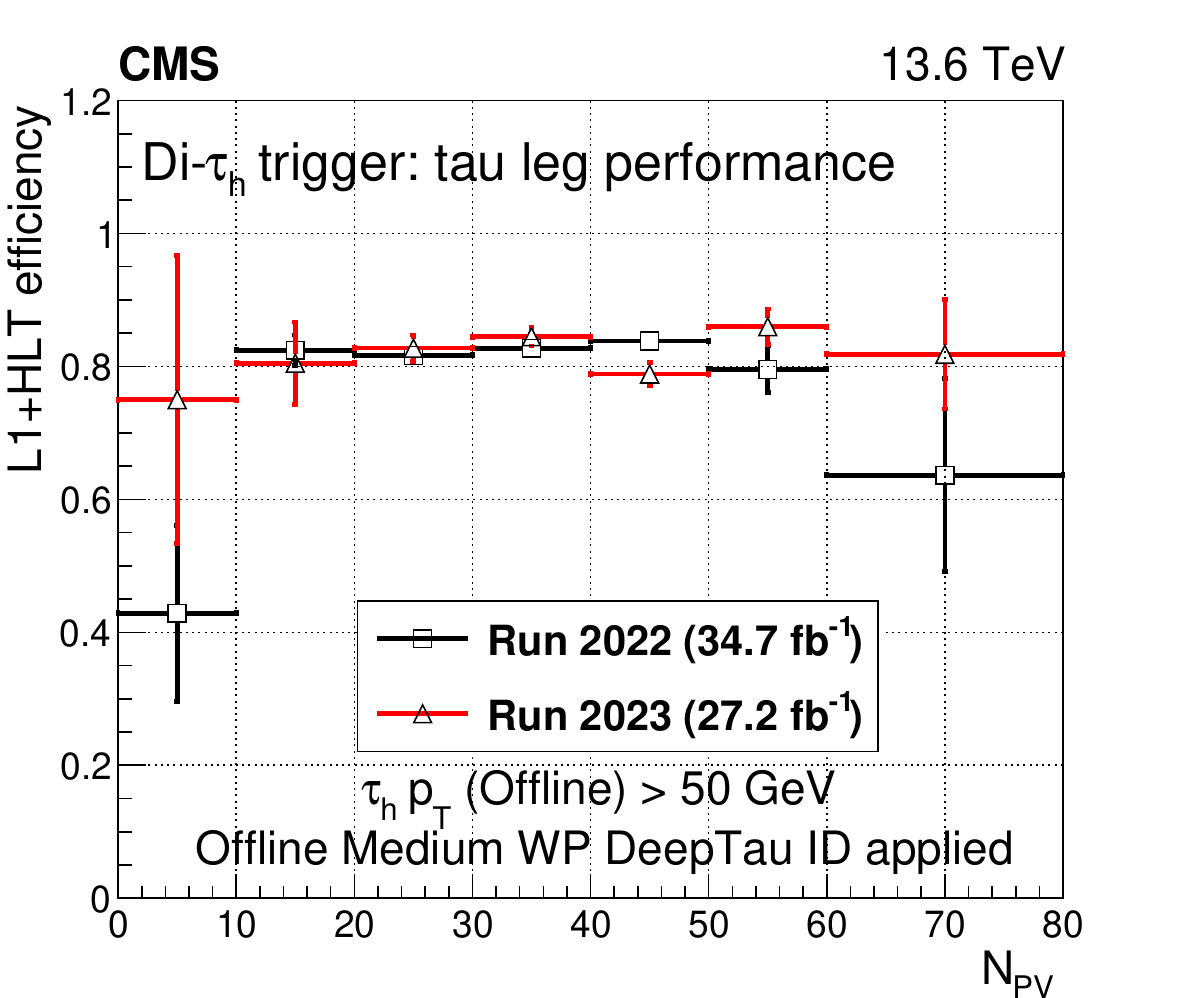}}
     \caption{A comparison of the L1\textplus HLT efficiency of the di-\tauh monitoring HLT path in 2022 and 2023 as a function of offline \tauh candidate \pt (upper left), \penEta (upper right), and \penPhi (lower left). The dependence on the \NPV is also shown (lower right). The uncertainties shown by the vertical bars are from the number of events available in the sample, while the horizontal bars show the bin width. Some of the vertical bars are smaller than the markers and are not shown.}\label{fig:ditau_2022_2023}
\end{figure*}

\begin{figure*}[hbtp!]
   \centering
     {\includegraphics[width=0.45\textwidth]{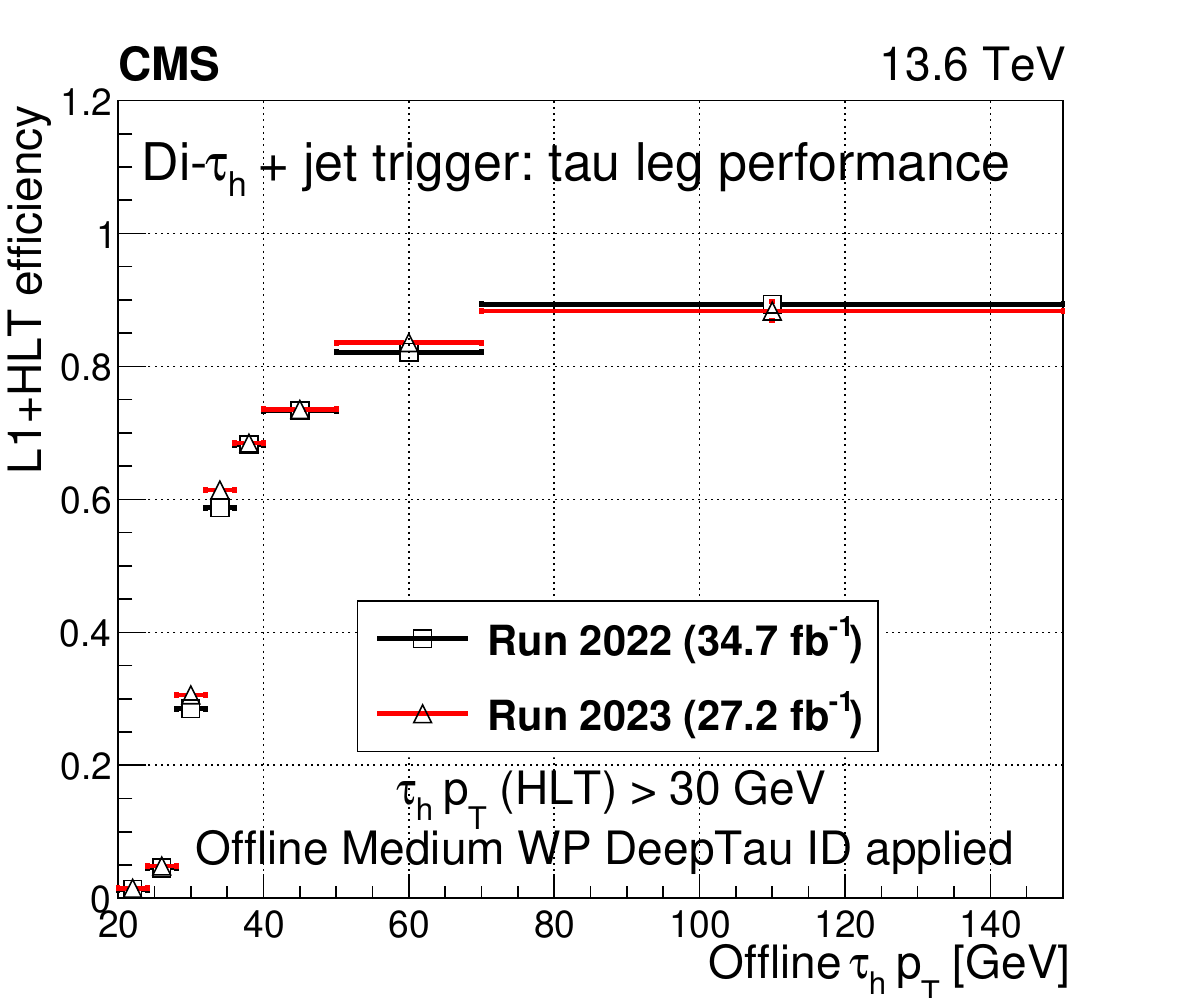}}
     {\includegraphics[width=0.45\textwidth]{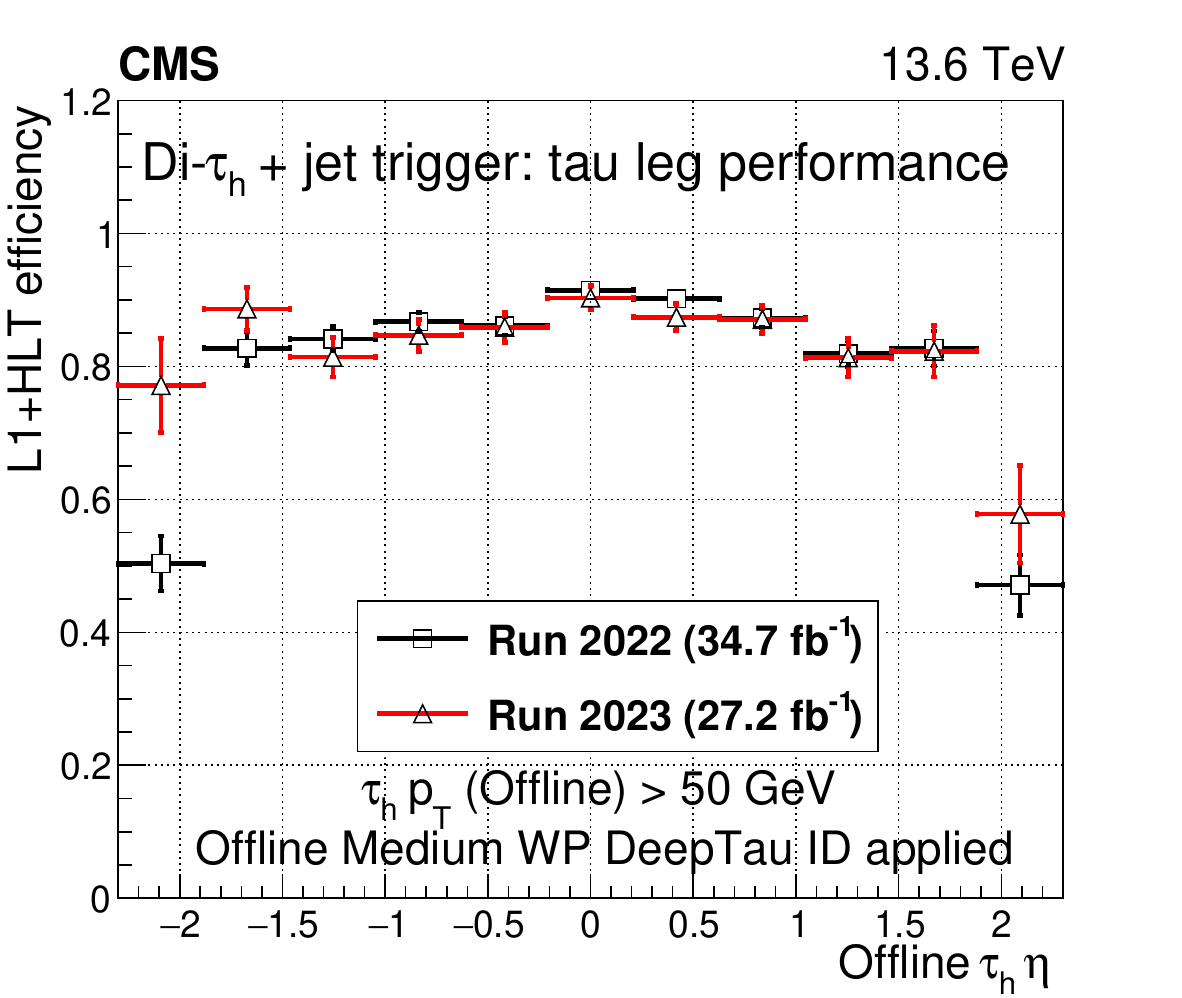}}
     {\includegraphics[width=0.45\textwidth]{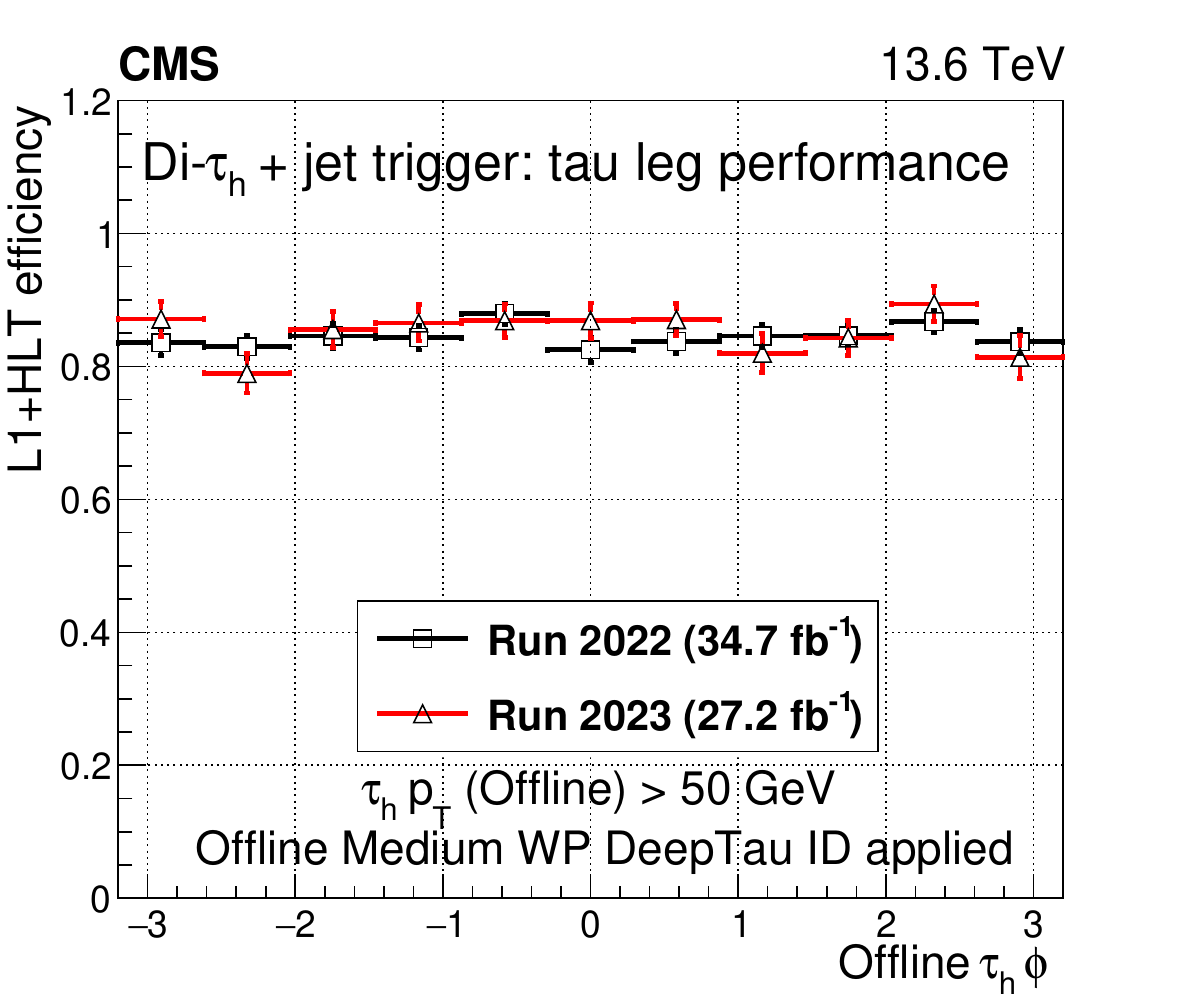}}
     {\includegraphics[width=0.45\textwidth]{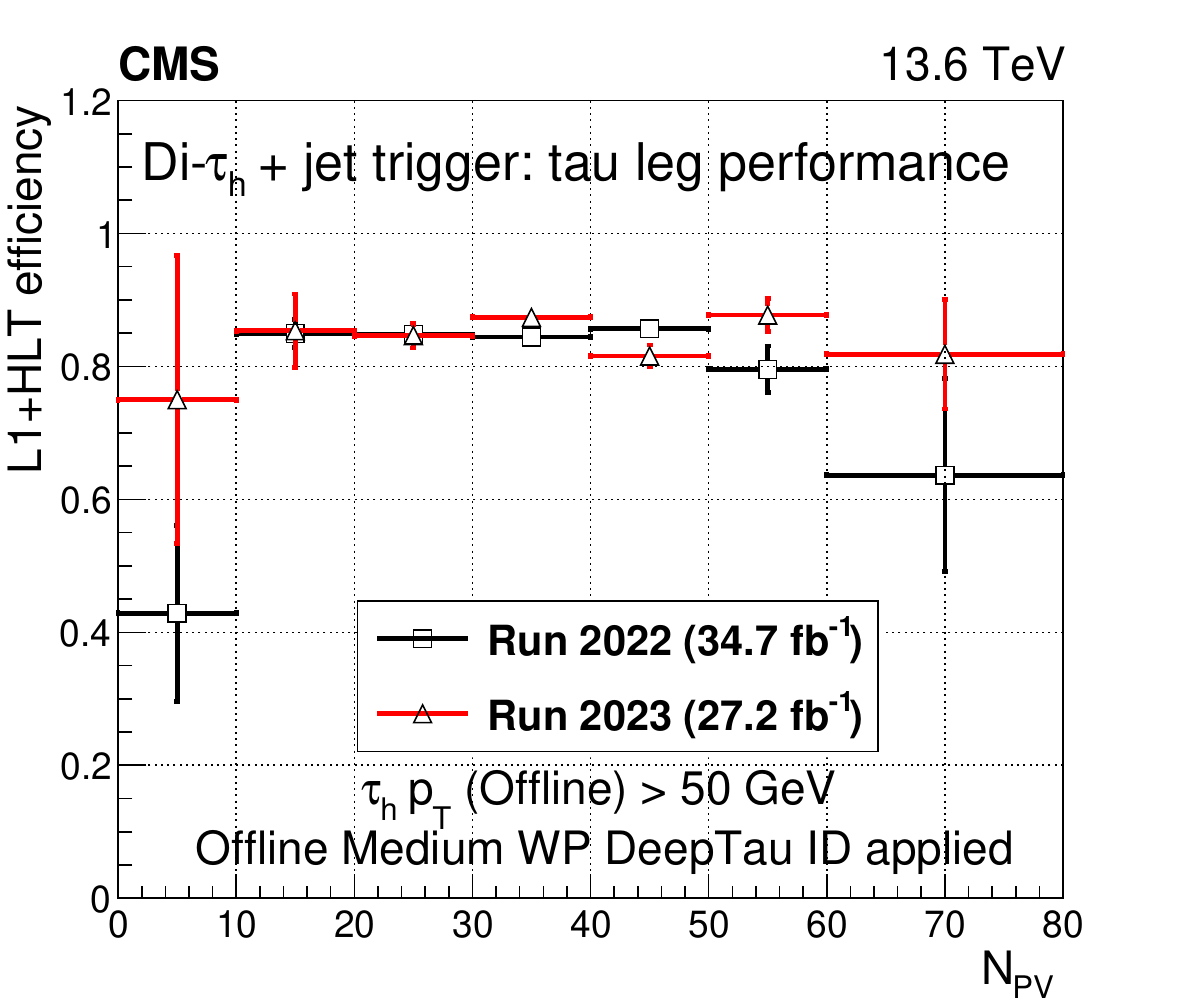}}
     \caption{A comparison of the L1\textplus HLT efficiency of the di-\tauhpj monitoring HLT path in 2022 and 2023 as a function of offline \tauh candidate \pt (upper left), \penEta (upper right), and \penPhi (lower left). The dependence on the \NPV is also shown (lower right). The uncertainties shown by the vertical bars are from the number of events available in the sample, while the horizontal bars show the bin width. Some of the vertical bars are smaller than the markers and are not shown.}\label{fig:ditaujet_2022_2023}
\end{figure*}

The efficiency of the \tauh leg in the di-\tauh HLT path is shown in Fig.~\ref{fig:ditau_2022_2023}, for 2022 and 2023. A modified version of the di-\tauh HLT path was produced for Run~3, called di-\tauhpj. It has online kinematic thresholds for \tauh candidates 5\GeV lower than the di-\tauh HLT path, and the additional requirement of an online jet with $\pt > 50\GeV$. This topology collects more low-\pt \tauh candidates than would be possible with the di-\tauh HLT path alone, at the cost of also requiring an energetic jet to reduce the HLT path's rate. The efficiency of the \tauh leg in this di-\tauhpj HLT path is shown in Fig.~\ref{fig:ditaujet_2022_2023}, for 2022 and 2023.
The HLT paths are evaluated in the same manner as the \etauh and \mutauh HLT paths previously, with the only differences being the higher online \pt thresholds of 35 (30)\GeV for the di-\tauh (di-\tauhpj) HLT path, and offline \tauh candidate \pt threshold of 50\GeV. Again, the HLT path is found to plateau, be robust against pileup, and have consistent performance between the two years of collected data.

\clearpage 

\clearpage 

\subsection{Efficiencies in simulation}
The efficiencies calculated with data from the previous subsection are now compared to efficiencies from simulation, using the same selection criteria as previously described per HLT path, and derived in the same DY enriched control region from the previous Section. The purpose of the comparison is to identify and measure differences between data and simulation, directly illustrated in a ratio of the efficiencies.
This ratio is called a scale factor, and is used together with its respective triggers to improve the accuracy of the detector response in simulation as compared with experiment. These scale factors are used in any analysis that uses the \tauh HLT paths. 

\begin{figure*}[hbtp!]
   \centering
     {\includegraphics[width=0.4\textwidth]{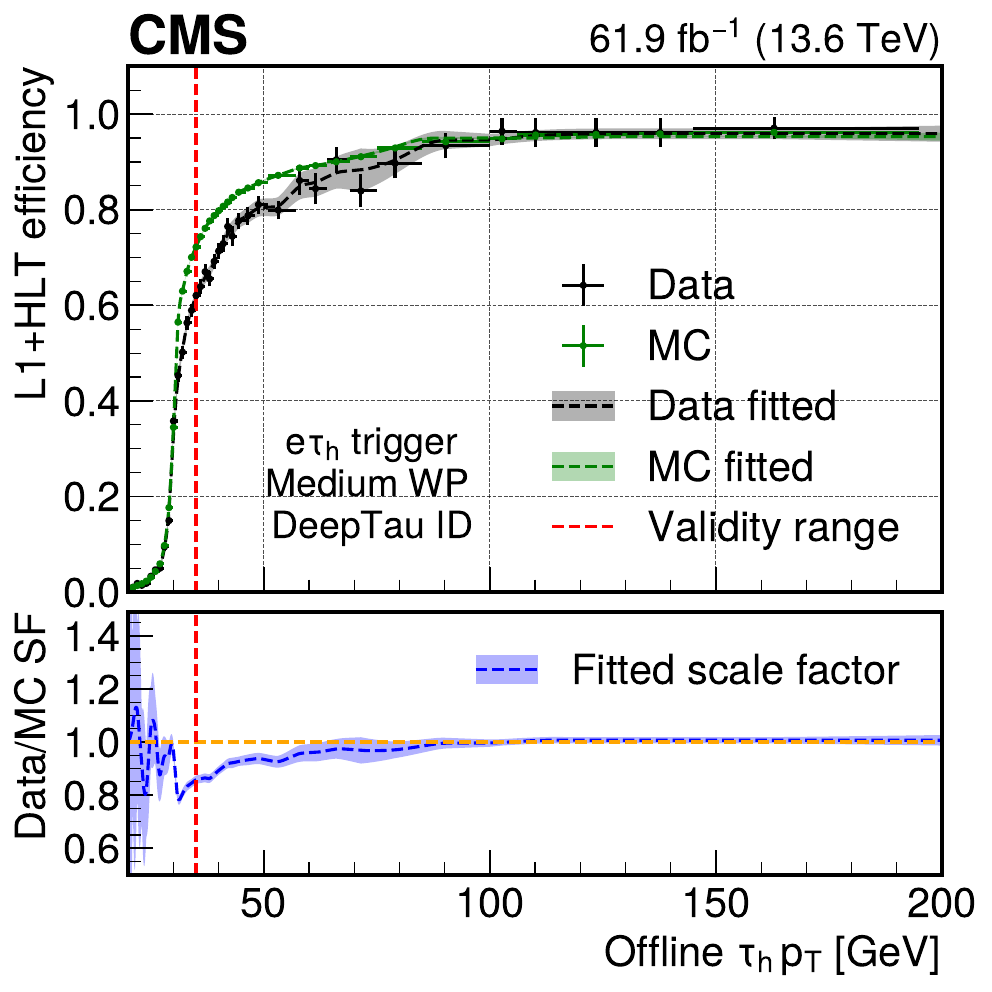}}
     {\includegraphics[width=0.4\textwidth]{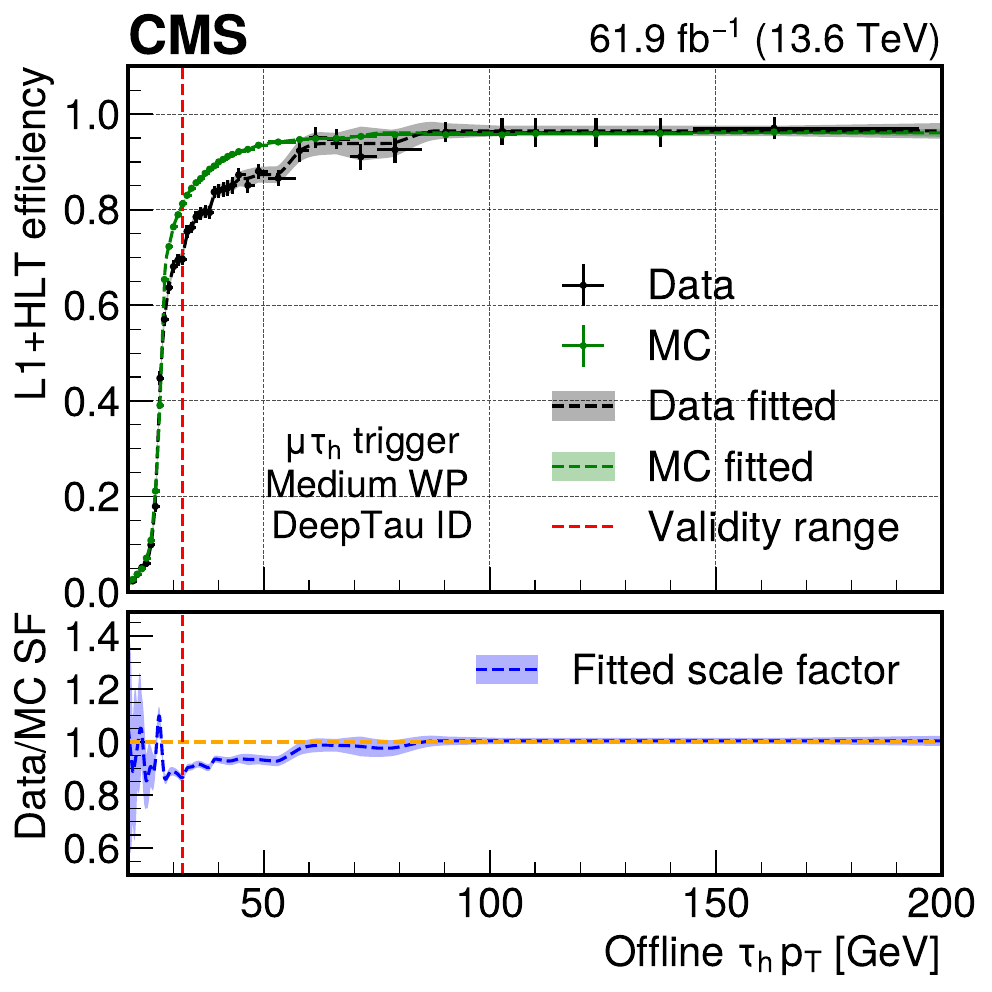}}
     {\includegraphics[width=0.4\textwidth]{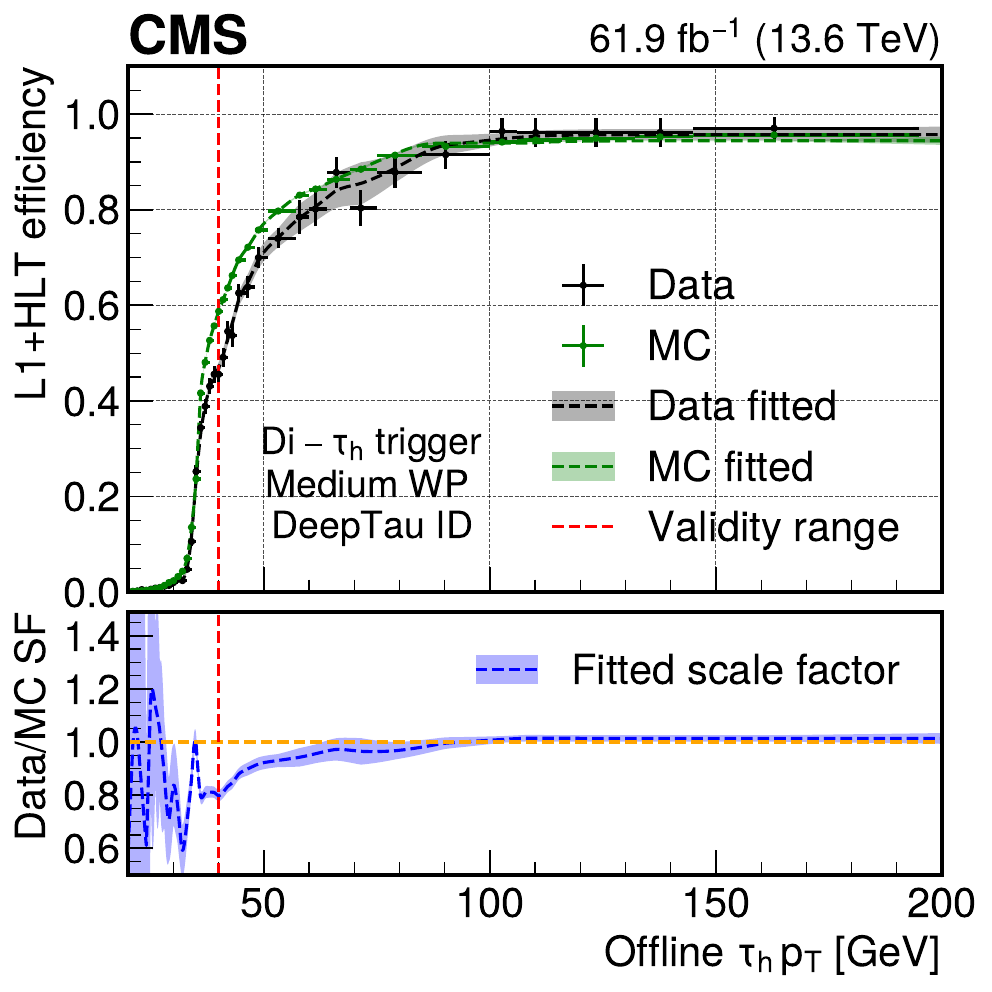}}
     {\includegraphics[width=0.4\textwidth]{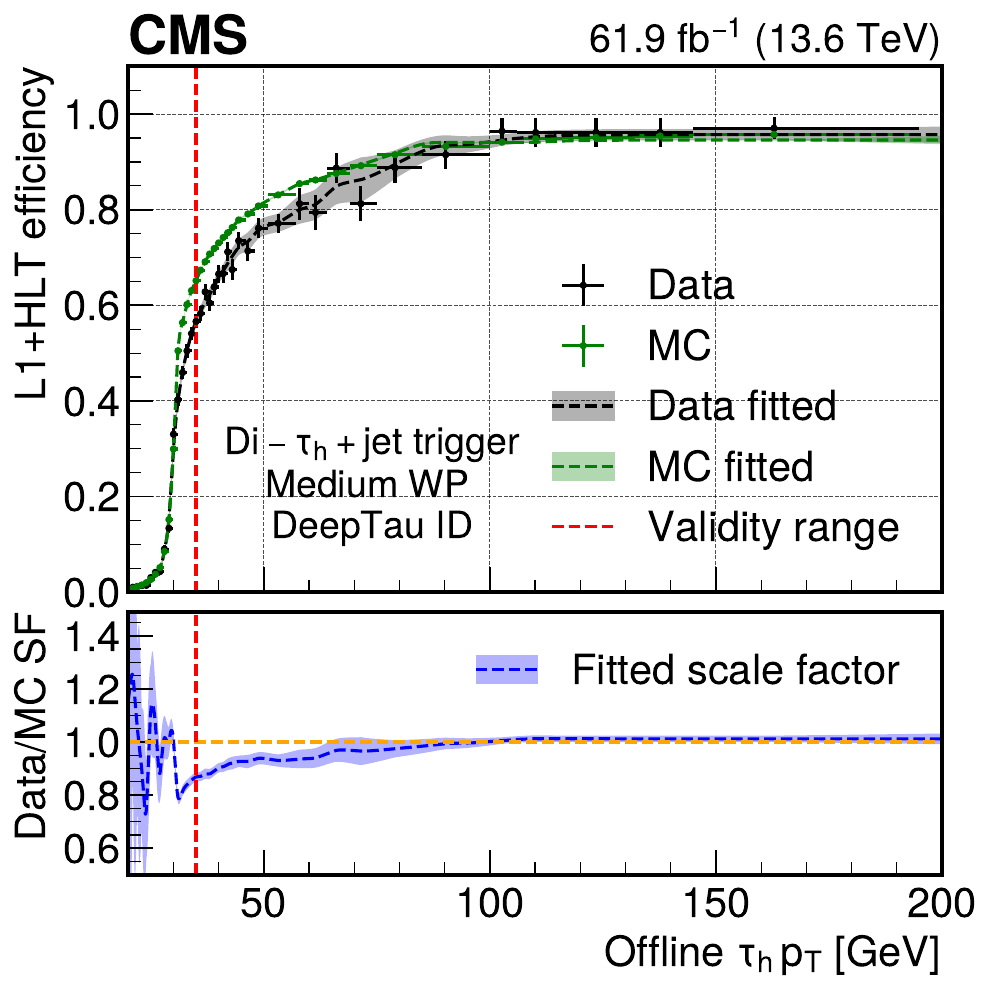}}
     \caption{Efficiencies and scale factors of the HLT monitoring paths using 2022--2023 data as functions of the offline \tauh candidate \pt for the \etauh (upper left), \mutauh (upper right), di-\tauh (lower left), and di-\tauhpj (lower right) HLT paths. The measured efficiencies for data are shown with black markers, and for simulation with green markers. The uncertainties shown by the vertical bars are from the number of events available in the sample, while the horizontal bars show the bin width. Some of the vertical bars are smaller than the markers and are not shown. The systematic uncertainties are negligible with respect to the statistical uncertainties and are not included in the bands. The corresponding dotted lines of the same color display the best fit results together with the statistical uncertainty bands. The scale factors, defined as the ratios of efficiencies between data fitted and simulation fitted from the upper panels, are shown in the lower panels as blue lines with associated blue uncertainty bands. Only values to the right of the red dotted line are used in physics studies, to avoid large statistical fluctuations, as they are sufficiently above the turn-on threshold of the trigger.}\label{fig:sf_allDMs_postEE}
\end{figure*}

The \tauh leg efficiencies and the corresponding scale factors of the \etauh, \mutauh, di-\tauh, and di-\tauhpj HLT paths are shown in Fig.~\ref{fig:sf_allDMs_postEE} for the 2022--2023 data combined. Here, the fitting is performed using a Gaussian process regressor. To avoid unreliable extrapolations, step functions are introduced to smoothly concatenate the low- and high-\pt regions. The dominant source of uncertainty is from the statistical uncertainties in data and simulation. The scale factors are found to be close to unity after the turn-on, indicating a good description of \tauh HLT paths in simulation.
The difference between data and simulation in the low-\pt region, below 60\GeV, is mainly due to the limited accuracy of HCAL energy response modeling, affecting the HCAL L1 trigger primitives in 2022--2023 simulations. 

\section{Summary}\label{sec:summary}
Two online machine-learning algorithms, \LtwoTauNNTag, described here for the first time, and \DeepTau, have been deployed into the high level trigger (HLT) to select hadronically decaying tau lepton (\tauh) candidates. Their performance has been evaluated using the data collected by the CMS experiment in proton-proton collisions at $\sqrt{s} = 13.6\TeV$ in 2022--2023, corresponding to an integrated luminosity of 62\fbinv. Comparisons to simulation were performed and show good agreement with the collected data, validating the current understanding of the HLT paths involving \tauh candidates. The updated HLT paths are found to deliver improved \tauh candidate identification efficiency without significantly increasing computational cost or event rate, allowing more genuine hadronic tau lepton decays to be collected at roughly the same resource cost as in 2018. These improvements benefit physics studies targeting final states with hadronically decaying tau leptons, including precision measurements of the Higgs boson, and searches beyond the standard model.

\begin{acknowledgments}
We congratulate our colleagues in the CERN accelerator departments for the excellent performance of the LHC and thank the technical and administrative staffs at CERN and at other CMS institutes for their contributions to the success of the CMS effort. In addition, we gratefully acknowledge the computing c,enters and personnel of the Worldwide LHC Computing Grid and other c,enters for delivering so effectively the computing infrastructure essential to our analyses. Finally, we acknowledge the enduring support for the construction and operation of the LHC, the CMS detector, and the supporting computing infrastructure provided by the following funding agencies: SC (Armenia), BMBWF and FWF (Austria); FNRS and FWO (Belgium); CNPq, CAPES, FAPERJ, FAPERGS, and FAPESP (Brazil); MES and BNSF (Bulgaria); CERN; CAS, MoST, and NSFC (China); MINCIENCIAS (Colombia); MSES and CSF (Croatia); RIF (Cyprus); SENESCYT (Ecuador); ERC PRG and PSG, TARISTU24-TK10 and MoER TK202 (Estonia); Academy of Finland, MEC, and HIP (Finland); CEA and CNRS/IN2P3 (France); SRNSF (Georgia); BMFTR, DFG, and HGF (Germany); GSRI (Greece); NKFIH (Hungary); DAE and DST (India); IPM (Iran); SFI (Ireland); INFN (Italy); MSIT and NRF (Republic of Korea); MES (Latvia); LMTLT (Lithuania); MOE and UM (Malaysia); BUAP, CINVESTAV, CONACYT, LNS, SEP, and UASLP-FAI (Mexico); MOS (Montenegro); MBIE (New Zealand); PAEC (Pakistan); MES, NSC, and NAWA (Poland); FCT (Portugal); MESTD (Serbia); MICIU/AEI and PCTI (Spain); MOSTR (Sri Lanka); Swiss Funding Agencies (Switzerland); MST (Taipei); MHESI (Thailand); TUBITAK and TENMAK (T\"{u}rkiye); NASU (Ukraine); STFC (United Kingdom); DOE and NSF (USA).

\hyphenation{Rachada-pisek} Individuals have received support from the Marie-Curie program and the European Research Council and Horizon 2020 Grant, contract Nos.\ 675440, 724704, 752730, 758316, 765710, 824093, 101115353, 101002207, 101001205, and COST Action CA16108 (European Union); the Leventis Foundation; the Alfred P.\ Sloan Foundation; the Alexander von Humboldt Foundation; the Science Committee, project no. 22rl-037 (Armenia); the Fonds pour la Formation \`a la Recherche dans l'Industrie et dans l'Agriculture (FRIA) and Fonds voor Wetenschappelijk Onderzoek contract No. 1228724N (Belgium); the Beijing Municipal Science \& Technology Commission, No. Z191100007219010, the Fundamental Research Funds for the Central Universities, the Ministry of Science and Technology of China under Grant No. 2023YFA1605804, the Natural Science Foundation of China under Grant No. 12535004, and USTC Research Funds of the Double First-Class Initiative No.\ YD2030002017 (China); the Ministry of Education, Youth and Sports (MEYS) of the Czech Republic; the Shota Rustaveli National Science Foundation, grant FR-22-985 (Georgia); the Deutsche Forschungsgemeinschaft (DFG), among others, under Germany's Excellence Strategy -- EXC 2121 ``Quantum Universe" -- 390833306, and under project number 400140256 - GRK2497; the Hellenic Foundation for Research and Innovation (HFRI), Project Number 2288 (Greece); the Hungarian Academy of Sciences, the New National Excellence Program - \'UNKP, the NKFIH research grants K 131991, K 133046, K 138136, K 143460, K 143477, K 146913, K 146914, K 147048, 2020-2.2.1-ED-2021-00181, TKP2021-NKTA-64, and 2025-1.1.5-NEMZ\_KI-2025-00004 (Hungary); the Council of Science and Industrial Research, India; ICSC -- National Research C,enter for High Performance Computing, Big Data and Quantum Computing, FAIR -- Future Artificial Intelligence Research, and CUP I53D23001070006 (Mission 4 Component 1), funded by the NextGenerationEU program (Italy); the Latvian Council of Science; the Ministry of Education and Science, project no. 2022/WK/14, and the National Science Center, contracts Opus 2021/41/B/ST2/01369, 2021/43/B/ST2/01552, 2023/49/B/ST2/03273, and the NAWA contract BPN/PPO/2021/1/00011 (Poland); the Funda\c{c}\~ao para a Ci\^encia e a Tecnologia (Portugal); the National Priorities Research Program by Qatar National Research Fund; MICIU/AEI/10.13039/501100011033, ERDF/EU, ``European Union NextGenerationEU/PRTR", and Programa Severo Ochoa del Principado de Asturias (Spain); the Chulalongkorn Academic into Its 2nd Century Project Advancement Project, the National Science, Research and Innovation Fund program IND\_FF\_68\_369\_2300\_097, and the Program Management Unit for Human Resources \& Institutional Development, Research and Innovation, grant B39G680009 (Thailand); the Eric \& Wendy Schmidt Fund for Strategic Innovation through the CERN Next Generation Triggers project under grant agreement number SIF-2023-004; the Kavli Foundation; the Nvidia Corporation; the SuperMicro Corporation; the Welch Foundation, contract C-1845; and the Weston Havens Foundation (USA).  
\end{acknowledgments}

\bibliography{auto_generated}

\cleardoublepage \appendix\section{The CMS Collaboration \label{app:collab}}\begin{sloppypar}\hyphenpenalty=5000\widowpenalty=500\clubpenalty=5000\input{TAU-24-002-public-authorlist.tex}\end{sloppypar}
\end{document}

%% file: TAU-24-002-public-authorlist.tex
\cmsinstitute{Yerevan Physics Institute, Yerevan, Armenia}
{\tolerance=6000
A.~Hayrapetyan, V.~Makarenko\cmsorcid{0000-0002-8406-8605}, A.~Tumasyan\cmsAuthorMark{1}\cmsorcid{0009-0000-0684-6742}
\par}
\cmsinstitute{Institut f\"{u}r Hochenergiephysik, Vienna, Austria}
{\tolerance=6000
W.~Adam\cmsorcid{0000-0001-9099-4341}, L.~Benato\cmsorcid{0000-0001-5135-7489}, T.~Bergauer\cmsorcid{0000-0002-5786-0293}, M.~Dragicevic\cmsorcid{0000-0003-1967-6783}, P.S.~Hussain\cmsorcid{0000-0002-4825-5278}, M.~Jeitler\cmsAuthorMark{2}\cmsorcid{0000-0002-5141-9560}, N.~Krammer\cmsorcid{0000-0002-0548-0985}, A.~Li\cmsorcid{0000-0002-4547-116X}, D.~Liko\cmsorcid{0000-0002-3380-473X}, M.~Matthewman, J.~Schieck\cmsAuthorMark{2}\cmsorcid{0000-0002-1058-8093}, R.~Sch\"{o}fbeck\cmsAuthorMark{2}\cmsorcid{0000-0002-2332-8784}, M.~Shooshtari\cmsorcid{0009-0004-8882-4887}, M.~Sonawane\cmsorcid{0000-0003-0510-7010}, W.~Waltenberger\cmsorcid{0000-0002-6215-7228}, C.-E.~Wulz\cmsAuthorMark{2}\cmsorcid{0000-0001-9226-5812}
% ...
\par}
\vskip\cmsinstskip
\dag:~Deceased\\
$^{1}$Also at Yerevan State University, Yerevan, Armenia\\
$^{2}$Also at TU Wien, Vienna, Austria\\
% ...
